\documentclass[journal]{IEEEtran}
\makeatletter

\usepackage{blindtext}
\usepackage[T1]{fontenc}
\usepackage{textcomp}
\usepackage{xurl}
\usepackage{tabularx}
\usepackage{booktabs}
\usepackage{makecell}
\usepackage{amsmath}
\usepackage{amsthm}

\usepackage{amssymb}
\usepackage{bm}
\usepackage{mdwmath}
\usepackage{cases}
\usepackage[short, c2]{optidef}
\usepackage{graphicx}
\graphicspath{{Figure/PDF/}{Biography/PDF/}}
\usepackage[dvipsnames]{xcolor}
\definecolor{myblue}{rgb}{0,0.4980,1} % Azure
\definecolor{myred}{rgb}{0.8706,0.1608,0.0627} % Chinese red
\usepackage[caption=false,font=footnotesize,subrefformat=parens]{subfig}
\usepackage[nosort,noadjust]{cite}
\usepackage{url}
\usepackage{tabularray}
\UseTblrLibrary{counter}
\usepackage[linesnumbered,ruled]{algorithm2e}

\usepackage{hyperref}
\newcommand{\colorhypersetup}{\@ifpackageloaded{hyperref}{\hypersetup{%
	bookmarksopen=true,%
	bookmarksnumbered=true,%
	pdfpagemode={UseOutlines},%default
	pdfstartview={FitH},%
	colorlinks=true,%
	linkcolor={myred},%
	citecolor={orange}
}}{\empty}}
\newcommand{\blackhypersetup}{\@ifpackageloaded{hyperref}{\hypersetup{%
	bookmarksopen=true,%
	bookmarksnumbered=true,%
	pdfpagemode={UseOutlines},%default
	pdfstartview={FitH},%
	colorlinks=true,%
	allcolors={black}
}}{\empty}}
 \blackhypersetup

\usepackage{mfirstuc-english}
\MFUhyphentrue
\usepackage[version3,description,IEEEtran]{eacro}
\acsetup{%
    pages/display=all,%
    pages/seq/use=false,%
    pages/name=true,%
    pages/fill={\quad},%
     make-links=false%
}
\DeclareAcronym{csi}{
	short = CSI,
	long = channel state information}

\DeclareAcronym{pcdm}{
	short = PCDM,
	long = physics-conditional diffusion model}

\DeclareAcronym{ntn}{
	short = NTN,
	long = non-terrestrial networks}

\DeclareAcronym{gs}{
	short = GS,
	long = ground station}

\DeclareAcronym{los}{
	short = LoS,
	long = line of sight}

\DeclareAcronym{stgnn}{
	short = ST-GNN,
	long = spatiotemporal graph neural network}

\DeclareAcronym{api}{
short = API,
long = application programming interface}

\DeclareAcronym{http}{
short = HTTP,
long = HyperText transfer protocol}

\DeclareAcronym{bs}{
	short = BS,
	long = base station}
 \DeclareAcronym{mg}{
	short = MG,
	long = multicast group}
  \DeclareAcronym{dt}{
	short = DT,
	long = digital twin}
 \DeclareAcronym{lstm}{
	short = LSTM,
	long = longs short-term memory}
  \DeclareAcronym{rnn}{
	short = RNN,
	long = recurrent neural networks}
   \DeclareAcronym{ddqn}{
	short = DDQN,
	long = double deep Q-network}
 \DeclareAcronym{rmse}{
	short = RMSE,
	long = root mean square error}
 \DeclareAcronym{svc}{
	short = SVC,
	long = scalable video coding}
 \DeclareAcronym{msvs}{
	short = MSVS,
	long = multicast short video streaming}
\DeclareAcronym{td3}{
	short = TD3,
	long =  twin delayed deep deterministic policy gradient}
 \DeclareAcronym{mdp}{
	short = MDP,
	long =  Markov decision process}
 \DeclareAcronym{dt3}{
	short = DFFTD3,
	long =  diffusion-based TD3}
\DeclareAcronym{smg}{
	short = SMG,
	long =  sub-MG}
 \DeclareAcronym{drl}{
	short = DRL,
	long =  deep reinforcement learning}
 \DeclareAcronym{nc}{
	short = NC,
	long =  network controller}
  \DeclareAcronym{qoe}{
	short = QoE,
	long =  quality of experience}
 \DeclareAcronym{dtp}{
	short = DTP,
	long =  DT data processing}
 \DeclareAcronym{vt}{
	short = VT,
	long =  video transcoding}
 \DeclareAcronym{mt}{
	short = MT,
	long =  multicast transmission}
\DeclareAcronym{ssim}{
	short = SSIM,
	long =  structural similarity index measure}
 \DeclareAcronym{dtu}{
	short = DTU,
	long =  DT model update}
  \DeclareAcronym{mse}{
	short = MSE,
	long =  mean square error}
  \DeclareAcronym{3C}{
	short = 3C,
	long =  communication computing and buffer control}
 
 \DeclareAcronym{noma}{
	short = NOMA,
	long =  non-orthogonal multiple access}
\DeclareAcronym{mcs}{
	short = MCS,
	long =  modulation coding scheme}

 \DeclareAcronym{ran}{
	short = RAN,
	long =  radio access network}

  \DeclareAcronym{PQoE}{
	short = PQoE,
	long =  personalized QoE}

   \DeclareAcronym{cdf}{
	short = CDF,
	long =  cumulative distribution function}

    \DeclareAcronym{gai}{
	short = GAI,
	long =  generative AI}
   \DeclareAcronym{udt}{
	short = UDT,
	long =  user DT}

 \DeclareAcronym{SNR}{
	short = SNR,
	long = signal-to-noise ratio}
 \DeclareAcronym{ap}{
	short = AP,
	long = access point}

     \DeclareAcronym{ffn}{
	short = FFN,
	long =feed-forward network}
    \DeclareAcronym{flop}{
	short = FLOP,
	long =floating point operation}

\newcommand{\upperroman}[1]{\uppercase\expandafter{\romannumeral#1}}

\newcommand{\myunit}[1]{%
	\ifmmode
		\text{#1}
	\else
		$ \text{#1} $% <-this % stops a space
	\fi}
\newcommand{\murm}{%
	\ifmmode
		\text{\textmu}
	\else
		\textmu
	\fi}

\newcommand{\MYnewpage}{%
	\ifCLASSOPTIONonecolumn
		\ifCLASSOPTIONjournal
			\typeout{The onecolumn journal mode.}
			\newpage
		\fi
	\fi}

\newlength{\mysinglefigwidth}
\newlength{\mymultifigwidth}
\ifCLASSOPTIONonecolumn
	\AtBeginDocument{\setlength{\mysinglefigwidth}{0.7\linewidth}}
\else
	\AtBeginDocument{\setlength{\mysinglefigwidth}{\linewidth}}
\fi
\ifCLASSOPTIONonecolumn
	\AtBeginDocument{\setlength{\mymultifigwidth}{0.5\linewidth}}
\else
	\AtBeginDocument{\setlength{\mymultifigwidth}{\linewidth}}
\fi

\makeatother
\begin{document}
%% *************************************************************************
\ifCLASSOPTIONonecolumn
    \typeout{The onecolumn mode.}
    \title{Physics-Informed Digital Twins for Channel Estimation and Traffic Prediction of Non-Terrestrial Networks}
    %\title{GenAI-Enabled Digital Twins for 6G NTN: Physics-Informed Channel Estimation and Traffic Prediction}
    \author{Xinyu~Huang,~\IEEEmembership{Member,~IEEE}, Yixiao~Zhang, Xue Qin, Mingcheng~He,~\IEEEmembership{Member,~IEEE}, Junling~Li,~\IEEEmembership{Member,~IEEE}, Weihua~Zhuang,~\IEEEmembership{Fellow,~IEEE}, and~Xuemin~(Sherman)~Shen,~\IEEEmembership{Fellow,~IEEE}
    
    \thanks{Xinyu Huang, Yixiao Zhang, Xue Qin, Mingcheng He, Weihua Zhuang, and Xuemin (Sherman) Shen are with the Department of Electrical and Computer Engineering, University of Waterloo, Waterloo, ON N2L 3G1, Canada (E-mail: \{x357huan, y3549zha, x7qin, m64he, wzhuang, sshen\}@uwaterloo.ca).}
    
    \thanks{Junling Li is with the National Mobile Communications Research Laboratory, School of Information Science and Engineering, Southeast University, Nanjing 211189, China (Email: junlingli@seu.edu.cn).}}

\else
    \typeout{The twocolumn mode.}
    %\title{3C Management for Adaptive Digital Twin Operation in QoE-Driven Multicast Short Videos}
    %\title{Fast Distributed LLM Inference Over Network Edges: A Digital Agent-Assisted Approach}
    %\title{Non-Terrestrial Network Digital Twin Construction for Channel Estimation and Traffic Prediction: A Physical-AI Approach}
    \title{Physics-Informed Digital Twins for Channel Estimation and Traffic Prediction of Non-Terrestrial Networks}
    %\title{GenAI-Enabled Digital Twins for 6G NTN: Physics-Informed Channel Estimation and Traffic Prediction}
    \author{Xinyu~Huang,~\IEEEmembership{Member,~IEEE}, Yixiao~Zhang, Xue Qin, Mingcheng~He,~\IEEEmembership{Member,~IEEE}, Junling~Li,~\IEEEmembership{Member,~IEEE}, Weihua~Zhuang,~\IEEEmembership{Fellow,~IEEE}, and~Xuemin~(Sherman)~Shen,~\IEEEmembership{Fellow,~IEEE}
    
    \thanks{Xinyu Huang, Yixiao Zhang, Xue Qin, Mingcheng He, Weihua Zhuang, and Xuemin (Sherman) Shen are with the Department of Electrical and Computer Engineering, University of Waterloo, Waterloo, ON N2L 3G1, Canada (E-mail: \{x357huan, y3549zha, x7qin, m64he, wzhuang, sshen\}@uwaterloo.ca).}
    
    \thanks{Junling Li is with the National Mobile Communications Research Laboratory, School of Information Science and Engineering, Southeast University, Nanjing 211189, China (Email: junlingli@seu.edu.cn).}}
   
\fi

\ifCLASSOPTIONonecolumn
	\typeout{The onecolumn mode.}
\else
	\typeout{The twocolumn mode.}
	%\markboth{IEEE Transactions on Wireless Communications}{Author \MakeLowercase{\textit{et al.}}: Title}
\fi

\maketitle

\ifCLASSOPTIONonecolumn
	\typeout{The onecolumn mode.}
	\vspace*{-50pt}
\else
	\typeout{The twocolumn mode.}
\fi
\begin{abstract}
In non-terrestrial networks (NTN), high-speed satellite orbital motion, limited pilot signaling resources, and spatiotemporally heterogeneous traffic make accurate channel and traffic state characterization particularly challenging. In this paper, we propose a physics-informed digital twin (DT) framework for channel estimation and traffic prediction. Particularly, it formulates channel state information (CSI) reconstruction as a controllable generative process guided by physical-prior tensors. Through a physics-aware attention mechanism, it effectively reconstructs the real-time full-resolution CSI from highly sparse and outdated pilots. Then, we develop an orbit-adaptive spatiotemporal graph neural network for traffic prediction. By leveraging a dual-stream attention mechanism to capture intra- and inter-plane spatial dependencies and a gated recurrent unit to model temporal evolution, the neural network effectively predicts stochastic traffic residuals, which are integrated with the deterministic physical traffic baseline to form the complete traffic state. To evaluate the proposed DT framework, we establish a high-fidelity NTN DT simulation platform based on real-world Starlink ephemeris, global population, and ERA5 weather data. Experimental results demonstrate that our framework significantly outperforms state-of-the-art baselines in both CSI reconstruction and traffic prediction accuracy.

%, achieving a 14.7\% reduction in traffic prediction error and decreasing CSI phase estimation error by over 51\%.

\end{abstract}

\ifCLASSOPTIONonecolumn
	\typeout{The onecolumn mode.}
	\vspace*{-10pt}
\else
	\typeout{The twocolumn mode.}
\fi

\IEEEpeerreviewmaketitle

\MYnewpage

%\section*{Acronyms}
%\acuseall
% \setlength{\mylabelwidth}{0.2\columnwidth}
%\IEEEprintacronyms
% \tableofcontents
%% *************************************************************************

\section{Introduction}
\label{sec:Introduction}

With the freezing of 3rd generation partnership project (3GPP) Release 17 and the ongoing evolution of Release 18/19, \ac{ntn} have been deemed as an indispensable component of the sixth-generation (6G) communication systems~\cite{9861699,cheng20226g,3gpp_ntn_overview}. By weaving thousands of low-earth orbit (LEO) satellites into a three-dimensional coverage fabric, NTN is expected to break geographical barriers and provide ubiquitous broadband access to unserved and underserved regions globally~\cite{10689514}. However, this paradigm shift, transitioning from static terrestrial base stations to high-speed orbiting celestial bodies, imposes unprecedented challenges on network management. The fundamental challenge lies in the extreme mobility of LEO satellites, which travel at velocities exceeding $7.5~\text{km/s}$, resulting in a network environment with high dynamics and spatiotemporal non-stationarity.

The effective management of such dynamic networks heavily relies on accurate digital twins (DTs), i.e., virtual representations that mirror the physical network states to enable proactive optimization~\cite{Liu2025,10148936,10605806,10818633}. However, constructing high-fidelity DTs for NTN faces two fundamental challenges: channel non-stationarity with sparse observations and spatiotemporal heterogeneity in traffic.
First, regarding the physical link, the fast satellite orbital motion induces rapid time-varying Doppler shifts and phase rotations. As satellite signals traverse the troposphere and ionosphere, they are affected by stochastic impairments such as rain attenuation and scintillation~\cite{10771645}, which induce rapid channel fluctuations. Combined with high satellite mobility, these dynamics make the channel highly time-varying, thus \ac{csi} estimated from pilot signals can become outdated even after a millisecond-level transmission delay. Additionally, due to the massive signaling overhead required for vast satellite footprints and strict transmission power limitations, allocating dense pilot signaling resources is forbidden, leaving the receiver with only highly sparse observations~\cite{10396843}. Traditional interpolation methods struggle to reconstruct high-fidelity instantaneous channels from these sparse and delayed samples, resulting in a significant degradation of beamforming gains. Second, regarding satellite traffic, LEO satellites continuously traverse diverse geographical environments, transitioning from oceanic to dense urban regions within minutes. This creates a unique motion-behavior coupling phenomenon: observed traffic fluctuations originate not only from the randomness of user behaviors but also from the deterministic geometric relationship between the satellite orbit and terrestrial population distribution. Conventional time-series forecasting models (e.g., long short-term memory (LSTM)~\cite{graves2012long} or standard Transformers~\cite{vaswani2017attention}) struggle to distinguish between user behavior changes and topology-induced traffic shifts, leading to poor generalization on unseen satellite orbital trajectories.

Existing frameworks to addressing these estimation and prediction problems typically fall into two categories. Model-based frameworks rely on predefined mathematical structures (e.g., geometry-based channel models or Poisson traffic processes)~\cite{dong2025uplink,11026808}. While physically interpretable, they are often too idealized to capture the intricate stochasticity of real-world environments, such as local weather dynamics or bursty user hotspots. Conversely, data-driven frameworks utilizing deep learning (DL) models treat the satellite network as a black box, which learn mapping functions solely from historical data~\cite{10970731,10938839}. Although effective, these frameworks often fail to focus on the extra domain knowledge in satellite systems, particularly the deterministic physical laws defined by Keplerian orbital mechanics and Maxwell’s equations. Neglecting these priors makes pure DL models highly data-hungry and prone to overfitting, thus limiting their reliability in the highly dynamic and data-sparse LEO environment.

To bridge this gap, we propose a physics-informed DT framework. The core idea is that LEO network dynamics are not purely random but are governed by a deterministic physical skeleton (e.g., satellite trajectory and free-space path loss (FSPL)) superimposed with stochastic environmental residuals (e.g., multipath scattering and user behaviors). Therefore, an efficient DT framework should not learn from scratch but instead leverage advanced artificial intelligence (AI) to incorporate deterministic physical laws as inductive biases into the learning process. This allows the neural network to focus on learning complex and nonlinear residuals that analytical models cannot capture.

The main contributions of this paper are summarized as follows:

\begin{itemize}
    \item Physics-informed DT framework: We propose a unified physics-informed DT framework for channel estimation and traffic prediction in NTN. By integrating deterministic physical priors with generative deep learning, we establish a new paradigm for state estimation in NTN.
    %, which offers superior generalization compared to purely data-driven baselines.

    \item Physics-conditional channel estimation: We develop a \ac{pcdm} for the channel DT. Unlike standard denoising models, it learns on physical prior tensors (e.g., satellite orbital dynamics and atmospheric conditions) through a physics-aware attention mechanism, which effectively solves the ill-posed inverse problem of recovering the real-time full-resolution CSI from sparse and outdated pilots.

    \item Orbit-adaptive traffic prediction: We develop an orbit-adaptive spatiotemporal graph neural network (ST-GNN) for the traffic DT. By isolating deterministic satellite orbital kinematics from stochastic user dynamics, the mobility-induced non-stationarity is effectively mitigated. Furthermore, a dual-stream graph attention mechanism in the neural network is proposed to capture intra- and inter-plane dependencies for precise traffic reconstruction.

    \item High-fidelity simulation: We establish a comprehensive DT simulation platform for NTN by integrating real-world Starlink trajectories, WorldPop demographic data, and ECMWF ERA5 weather data. Experimental results show that our framework significantly reduces prediction errors as compared to state-of-the-art benchmarks.
    
\end{itemize}

The remainder of this paper is organized as follows. Section~\ref{related} reviews related work. Section~\ref{cdt_sec} details the \ac{pcdm} for channel estimation. Section~\ref{tdt_sec} describes the orbit-adaptive ST-GNN for traffic prediction. Section~\ref{exp} presents the experimental results, and Section~\ref{sec:Conclusion} concludes this study.

\section{Related Work}\label{related}
In this section, we review existing work on three aspects: 6G NTN DTs, sparse channel estimation for DTs, and spatiotemporal traffic prediction for DTs.

\subsection{Digital Twins for 6G Non-Terrestrial Networks}
As a cornerstone of the 6G architecture, the DT paradigm is transitioning from static monitoring to proactive network management. In the context of NTN, a high-fidelity DT is envisioned to mirror the time-varying satellite orbital dynamics and electromagnetic environments in real-time and enable zero-touch network management~\cite{10345669,11006653}. Existing literature has explored DT deployment for satellite networks and presented foundational frameworks for topology visualization, resource orchestration, and network slicing based on statistical models and offline simulators~\cite{10318165,10742580}.
Building upon these foundational works, a key objective in advancing NTN DTs is achieving real-time synchronization to capture the instantaneous dynamics of LEO constellations. Recent research has increasingly adopted data-driven AI methodologies in high-mobility NTNs to handle domain-specific complexities~\cite{10938839,9439942}. At the physical layer, a primary focus is on reconstructing instantaneous channels under extreme data sparsity and rapid aging. At the network layer, accurate modeling requires differentiating between deterministic topology-driven traffic changes and stochastic user requests. Consequently, to realize efficient NTN DTs, there is a growing trend toward physics-informed AI, which integrates deterministic physical laws to guide neural network training and inference.

\subsection{Dynamic Sparse Channel Estimation for Digital Twins}
To achieve high fidelity in the physical layer, accurate CSI is a prerequisite. Traditional estimation methods, such as least squares (LS) and minimum mean square error (MMSE), provide foundational baselines for channel estimation. In LEO scenarios, relative velocities exceeding $7.5~\text{km/s}$ induce severe Doppler shifts, resulting in rapid channel aging~\cite{9586735}. Capturing these high-frequency temporal variations typically requires dense pilot signals. Given the strict radio resources and transmission power constraints in NTN, optimizing pilot overhead is challenging. To address this, compressed sensing techniques have been applied that exploit angular-domain sparsity to reduce signaling overhead~\cite{11026808}. Recently, DL has emerged as a powerful methodology for CSI estimation. Discriminative models (e.g., convolutional neural networks (CNNs) and recurrent neural networks (RNNs)) have been successfully utilized to map sparse pilots to full CSI~\cite{9439942,10637286}. To further capture the underlying probability distributions of wireless channels, generative models, including generative adversarial networks (GANs) and variational autoencoders (VAEs), have been introduced to reconstruct missing channel features~\cite{10589572,10944428}. These models demonstrate strong capabilities in capturing statistical characteristics by processing channel matrices through established image or time-series representations. Building upon these data-driven foundations, a promising recent trend focuses on embedding electromagnetic and geometric priors into neural networks to further enhance channel fidelity in NTN DTs.

\subsection{Spatiotemporal Traffic Prediction for Digital Twins}
Satellite traffic exhibits a unique motion-behavior coupling effect, where traffic fluctuations are driven by both stochastic user behaviors and deterministic satellite orbital movement. To capture temporal traffic patterns, early works have utilized classical time-series analysis and temporal neural networks~\cite{10570573,peng2025spatio}. To further model spatial dependencies, GNNs have been adopted from terrestrial networks. These approaches typically construct graphs based on inter-satellite links (ISLs) to aggregate spatial information across the constellation~\cite{10900453,10755127}. Since LEO satellites operate on a dynamic spherical manifold where neighbor relationships and relative positions change continuously, a key focus in recent research is extending these graph-based models to capture time-varying topologies. Consequently, modeling the interplay between motion-induced traffic shifts and actual user demand variations during cross-regional handovers has emerged as a primary direction for enhancing traffic prediction accuracy in NTN DTs.

\subsection{Distinctions of Our Work}
The primary distinction of our work lies in the integration of deterministic physical priors into generative AI-based DTs.
Different from purely data-driven models that learn from scratch, our proposed \ac{pcdm} utilizes known physical parameters, such as satellite trajectories and atmospheric attenuation, as deterministic conditions to construct the channel DT. This allows the model to focus on learning the complex environmental residuals rather than rediscovering fundamental physical laws.
In traffic prediction, unlike standard ST-GNNs, our orbit-adaptive ST-GNN decouples the traffic into a physics-determined baseline and a behavior-driven residual to construct the traffic DT. By leveraging a dual-attention mechanism to capture traffic dependencies, our proposed neural network can adapt well to the dynamic geometry of LEO constellations.

\section{Channel Digital Twin Construction}\label{cdt_sec}
In this section, we propose a \ac{pcdm} to construct a high-fidelity channel DT. We consider a discrete-time NTN system indexed by slot $n \in \mathbb{Z}^{+}$, where each slot has duration $\Delta_t$. Let $t_n = n\Delta_t$ denote the physical time corresponding to the initial instant of slot~$n$. The channel is assumed to be quasi-static within each slot, i.e., the CSI remains approximately constant over time interval $[t_n, t_n+\Delta_t)$. Under this assumption, the channel DT aims to infer the real-time full-resolution \ac{csi} map from delayed and extremely sparse pilot observations. 

We consider downlink \ac{ntn} overlaid on a geographic region discretized into an $N_x \times N_y$ spatial grid. The system operates over a radio spectrum bandwidth divided into $N_c$ orthogonal subcarriers. The ground-truth \ac{csi} at slot $n$ is represented by a 3D complex tensor $\pmb{\mathcal{H}}_n \in \mathbb{C}^{N_x \times N_y \times N_c}$, where $[\pmb{\mathcal{H}}_n]_{i,j,k}$ denotes the channel coefficient at the center of grid cell $(i,j)$ for the $k$-th subcarrier. For each grid cell, we consider only the dominant satellite-ground link at each slot.

%The objective is to predict the current full-resolution \ac{csi} map at transmission slot $t$, based on extremely sparse pilot observations acquired at past time slot $t-\tau$, where $\tau$ represents the temporal delay between pilot acquisition and the actual downlink transmission.

%We consider downlink \ac{ntn} overlaid on a geographic region discretized into an $N_x \times N_y$ spatial grid. The system operates over a bandwidth divided into $N_c$ orthogonal subcarriers. The ground-truth \ac{csi} at time slot $t$ is mathematically represented by a 3D complex tensor $\pmb{\mathcal{H}}_t \in \mathbb{C}^{N_x \times N_y \times N_c}$. The entry $[\pmb{\mathcal{H}}_t]_{i,j,k}$ denotes the channel coefficient at spatial coordinate $(i,j)$ for the $k$-th subcarrier.

Due to signaling latency and processing delay in \ac{ntn}, the CSI available at the ground station for slot $n$ is outdated. Let $\tau$ denote the end-to-end delay between pilot acquisition and its use for downlink transmission. This continuous-time delay is transformed to an integer slot delay, $d = \left\lceil {\tau}/{\Delta_t} \right\rceil$.
Therefore, the most recent available CSI observation for slot $n$ is the sparse pilot measurement collected at slot $n-d$, given by
\begin{equation}
    \pmb{\mathcal{Y}}_{n-d} = \pmb{\mathcal{M}}_{n-d} \odot \left(\pmb{\mathcal{H}}_{n-d} + \pmb{\mathcal{N}}_{n-d}\right),
\end{equation}
where $\pmb{\mathcal{Y}}_{n-d} \in \mathbb{C}^{N_x \times N_y \times N_c}$ is the sparse observation tensor, $\pmb{\mathcal{H}}_{n-d}$ is the historical CSI tensor, $\pmb{\mathcal{N}}_{n-d}$ denotes the additive Gaussian noise tensor, symbol $\odot$ is the Hadamard product, and $\pmb{\mathcal{M}}_{n-d} \in \{0,1\}^{N_x \times N_y \times N_c}$ is the binary sampling mask indicating pilot locations at slot $n-d$. Here, $[\pmb{\mathcal{M}}_{n-d}]_{i,j,k}=1$ means that a pilot symbol is observed at the center of grid cell $(i,j)$ on subcarrier $k$, and $[\pmb{\mathcal{M}}_{n-d}]_{i,j,k}=0$ otherwise. 

%Due to the signaling latency and processing delay in satellite networks, the available \ac{csi} at the \ac{gs} at current transmission slot $t$ is inevitably outdated. The available observation at time slot $t$ is a delayed measurement collected at past time slot $t-\tau$, given by
%\begin{equation}
 %   \pmb{\mathcal{Y}}_{t-\tau} = \pmb{\mathcal{M}}_{t-\tau} \odot \left(\pmb{\mathcal{H}}_{t-\tau} + \pmb{\mathcal{N}}_{t-\tau}\right),
%\end{equation}
%where $\pmb{\mathcal{Y}}_{t-\tau} \in \mathbb{C}^{N_x \times N_y \times N_c}$ is the sparse observation tensor, $\pmb{\mathcal{H}}_{t-\tau}$ represents the historical \ac{csi}, and $\pmb{\mathcal{M}}_{t-\tau} \in \{0, 1\}^{N_x \times N_y \times N_c}$ is the binary sampling mask tensor indicating the pilot locations at time slot $t-\tau$. $[\pmb{\mathcal{M}}_{t-\tau}]_{i,j,k} = 1$ indicates that a pilot symbol was transmitted at spatial location $(i,j)$ on subcarrier $k$ at time slot $t-\tau$, and $[\pmb{\mathcal{M}}_{t-\tau}]_{i,j,k} = 0$ otherwise. $\pmb{\mathcal{N}}_{t-\tau}$ is the additive Gaussian noise tensor. Symbol $\odot$ denotes the element-wise Hadamard product.

The fundamental challenge for the channel DT is to infer real-time CSI tensor $\pmb{\mathcal{H}}_n$ from delayed and sparse observation $\pmb{\mathcal{Y}}_{n-d}$. This is a spatiotemporal estimation problem that requires extrapolating the channel evolution across a delay of $d$ slots. To regularize this inference, we introduce a multi-channel\footnote{The term `channel' here refers to the tensor's feature map dimension, rather than the communication channel.} physics tensor $\pmb{\mathcal{P}}_n$, which encodes the deterministic physical state of the satellite-ground propagation environment at slot $n$. This physical prior guides the \ac{pcdm} to align the delayed observation with the real-time physical state to improve CSI reconstruction accuracy. Based on the delayed observation and the real-time physical prior, we propose to learn a conditional generative distribution $p_\theta(\pmb{\mathcal{H}}_n \mid \pmb{\mathcal{Y}}_{n-d}, \pmb{\mathcal{P}}_n)$.  %The detailed construction of $\pmb{\mathcal{P}}_n$ is presented next.

\subsection{Multi-Channel Physics Tensor}
To capture the deterministic physical factors affecting channel evolution, we construct a 6-channel physics tensor $\pmb{\mathcal{P}}_n\in\mathbb{R}^{N_x \times N_y \times 6}$ for slot $n$. It is obtained by sampling the continuous-time physical state at $t_n=n\Delta_t$, and serves as a deterministic prior for the channel amplitude and phase dynamics.
Let $\mathbf{p}_\text{s}(t_n)\in\mathbb{R}^3$ and $\mathbf{v}_\text{s}(t_n)\in\mathbb{R}^3$ denote the dominant satellite position and velocity vectors at time $t_n$, obtained from the simplified general perturbations-4 (SGP4) model in the earth-centered, earth-fixed (ECEF) coordinate system. Let $\mathbf{p}_\text{g}(i,j)\in\mathbb{R}^3$ be the ECEF coordinate of reference point associated with grid cell $(i,j)$, chosen as its center. The slant range between the satellite and the center of grid cell $(i,j)$ at time $t_n$ is given by $D_{i,j}(t_n)=\left\|\mathbf{p}_\text{s}(t_n)-\mathbf{p}_\text{g}(i,j)\right\|_2$.

The 6-channel physics tensor consists of three groups of deterministic priors: energy channels (Channels 1--3) for large-scale fading, a motion channel (Channel 4) for Doppler dynamics, and environment channels (Channels 5--6) for scattering-related statistics. Each channel is defined as follows.

\textit{Channel 1: Free-space path loss (dB).} According to the Friis transmission formula, the first channel is given by
\begin{equation}
    [\pmb{\mathcal{P}}_n]_{i,j,1}
    = 20\log_{10}\left(\frac{4\pi D_{i,j}(t_n) f_c}{c}\right),
\end{equation}
where $f_c$ is the carrier frequency and $c$ is the speed of light.

\textit{Channel 2: Satellite antenna gain (dBi).} This channel depends on the satellite attitude and beamforming strategy. Let $G_\text{tx}(\theta,\phi)$ denote the directional transmit antenna gain in dBi, where $\theta_{i,j}(t_n)$ and $\phi_{i,j}(t_n)$ are the azimuth and elevation angles of the center of grid cell $(i,j)$ relative to the satellite boresight at time $t_n$. Thus, the second channel is given by
\begin{equation}
    [\pmb{\mathcal{P}}_n]_{i,j,2}
    = G_\text{tx}\!\left(\theta_{i,j}(t_n), \phi_{i,j}(t_n)\right).
\end{equation}

\textit{Channel 3: Atmospheric attenuation (dB).} For Ku/Ka/Q/V bands, atmospheric attenuation is mainly dominated by rain attenuation. Thus, we model the third channel as
\begin{equation}
    [\pmb{\mathcal{P}}_n]_{i,j,3}
    = A_\mathrm{r}\!\left(R_{i,j}, \epsilon_{i,j}(t_n), f_c\right),
\end{equation}
where $\epsilon_{i,j}(t_n)$ is the elevation angle of the center of grid cell $(i,j)$ at time $t_n$, and $A_\mathrm{r}(\cdot)$ is obtained from the ITU-R P.618 recommendation~\cite{618} based on the local rainfall rate map $R_{i,j}$.

\textit{Channel 4: Doppler shift (Hz).} This channel captures the high-mobility effect of LEO satellites and provides a deterministic reference for phase evolution across the observation delay. Let
$\mathbf{u}_{i,j}(t_n)=({\mathbf{p}_\text{g}(i,j)-\mathbf{p}_\text{s}(t_n)})/{D_{i,j}(t_n)}$ denote the unit direction vector from the satellite to the center of grid cell $(i,j)$ at time $t_n$. Then, the fourth channel is given by
\begin{equation}
    [\pmb{\mathcal{P}}_n]_{i,j,4}
    =
    \frac{f_c}{c}
    \left[
    \mathbf{v}_\text{s}(t_n)^\mathrm{T}\mathbf{u}_{i,j}(t_n)
    \right].
\end{equation}

\textit{Channel 5: Land-cover semantic label.} Different terrains bring different multipath characteristics, e.g., Rician fading in rural areas and Rayleigh fading in dense urban areas. Let $\mathbb{M}_\mathrm{G}: \mathbb{R}^3 \rightarrow \Omega_\mathrm{c}$ denote a discrete mapping based on global land-cover databases (e.g., ESA WorldCover), where $\Omega_\mathrm{c}=\{0:\text{Ocean},1:\text{Rural},2:\text{Urban},\dots\}$. Then, the fifth channel is given by
\begin{equation}
    [\pmb{\mathcal{P}}_n]_{i,j,5}
    =
    \mathbb{M}_\mathrm{G}(\mathbf{p}_\text{g}(i,j)).
\end{equation}
%Although stored as integers, these labels are processed through an embedding layer to capture categorical relationships.

\textit{Channel 6: Tropospheric scintillation index.} For satellite links in Ku/Ka/Q/V bands, atmospheric turbulence causes rapid amplitude fluctuations, known as tropospheric scintillation. Its intensity is quantified by the $S_4$ index, given by
\begin{equation}
    S_4 \triangleq
    \sqrt{\frac{\langle I^2\rangle-\langle I\rangle^2}{\langle I\rangle^2}},
\end{equation}
where $I$ is the signal intensity and symbol $\langle \cdot \rangle$ denotes statistical expectation. Since the instantaneous $I$ is unavailable during channel DT construction, we use a deterministic predictor $\mathcal{F}_\mathrm{S}(\cdot)$ based on ITU-R P.618 recommendation to estimate the scintillation level. Therefore, the sixth channel is given by
\begin{equation}
    [\pmb{\mathcal{P}}_n]_{i,j,6}
    =
    \mathcal{F}_\mathrm{S}(\mathbf{p}_\text{g}(i,j), t_n, f_c).
\end{equation}

  %  Channel 5: Land cover semantic label. Different terrains bring distinct multipath distributions (e.g., Rician fading for rural areas vs. Rayleigh fading for dense urban environments). We define a discrete mapping function $\mathbb{M}_\text{GIS}: \mathbb{R}^3 \to \Omega_\mathrm{c}$ based on global land cover databases (e.g., ESA WorldCover), where $\Omega_\mathrm{c} = \{0: \text{Ocean}, 1: \text{Rural}, 2: \text{Urban}, \dots\}$. Channel 5 stores these semantic labels\footnote{Note that while stored as integers, these labels are processed via an embedding layer to capture categorical relationships.} as follows:
%    \begin{equation}
%        [\pmb{\mathcal{P}}_t]_{i,j,5} = \mathbb{M}_\text{GIS}(\mathbf{p}_\text{g}(i,j)).
%    \end{equation}

%It is worth noting that the current physics tensor $\pmb{\mathcal{P}}_t$ primarily captures macroscopic environmental dynamics (e.g., tropospheric scintillation and atmospheric attenuation). Incorporating extreme and low-probability space weather anomalies (e.g., severe solar flares) into the deterministic skeleton remains a limitation of the current model, as such events lack predictable temporal regularity.

\subsection{Physics-Conditional Diffusion Process}
\begin{figure}[!t]
    \centering
\includegraphics[width=\linewidth]{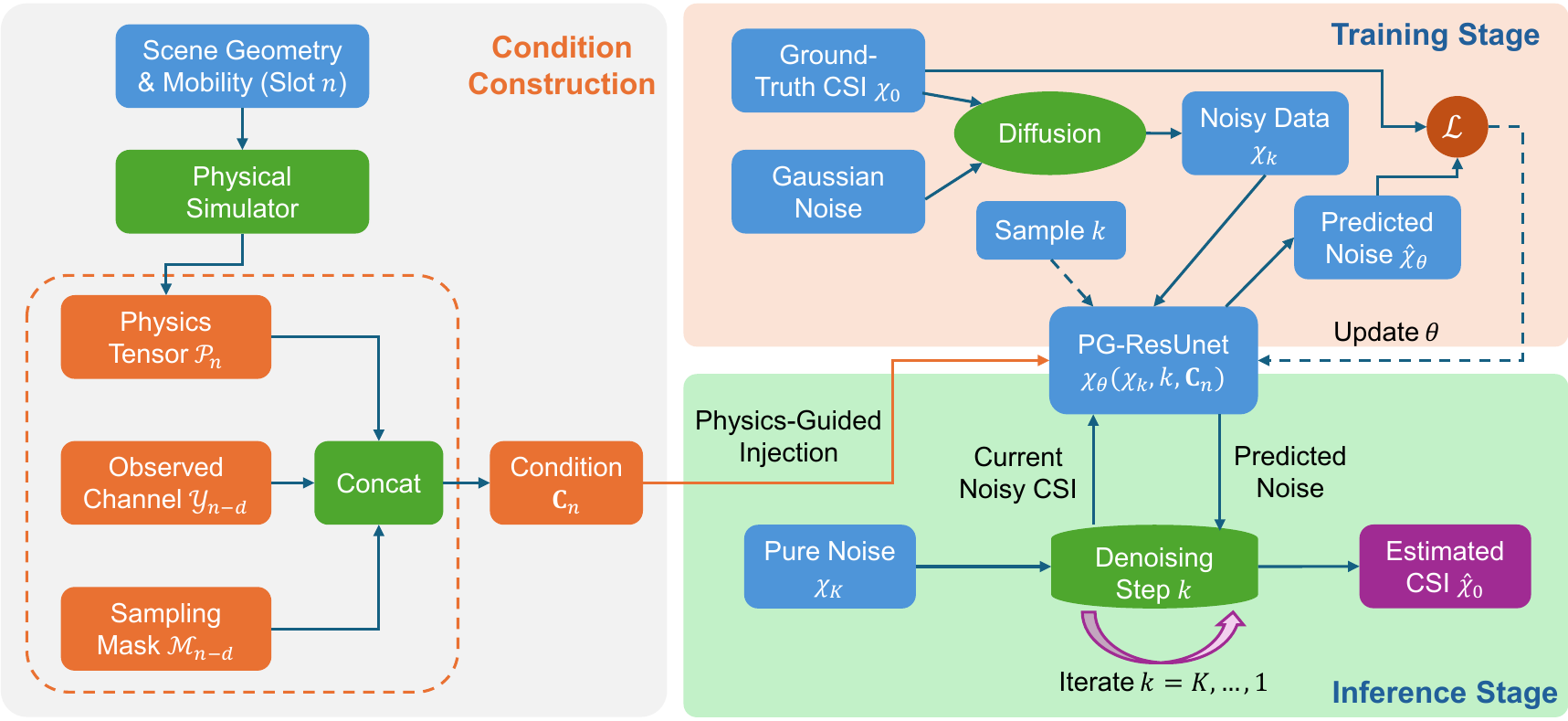}
    \caption{The proposed physics-conditional diffusion model architecture.}
    \label{pcdm}
\end{figure}

Given the deterministic physical prior, $\pmb{\mathcal{P}}_n$, we propose the \ac{pcdm} to reconstruct the CSI tensor at slot $n$. We treat complete channel tensor $\pmb{\mathcal{H}}_n$ as a sample from real data distribution $q(\pmb{\mathcal{H}}_n)$. The proposed \ac{pcdm} consists of a condition construction module, a forward diffusion process used in training, and a physics-guided reverse generative process used in inference, as shown in Fig.~\ref{pcdm}.

\subsubsection{Forward Diffusion Process}
The forward process is a fixed Markov chain that gradually adds Gaussian noise to the ground-truth channel tensor, $\pmb{\mathcal{X}}_0 \triangleq \pmb{\mathcal{H}}_n$, over $K$ discrete diffusion steps~\cite{ho2020denoising}.\footnote{A diffusion step is one iteration of the internal Markov diffusion process.} At diffusion step $k$, the transition distribution to noisy latent variable $\pmb{\mathcal{X}}_k$ is given by
\begin{equation}
    q(\pmb{\mathcal{X}}_k \mid \pmb{\mathcal{X}}_{k-1})
    =
    \mathcal{N}\!\left(
    \pmb{\mathcal{X}}_k;
    \sqrt{1-\beta_k}\,\pmb{\mathcal{X}}_{k-1},
    \beta_k \mathbf{I}
    \right),
\end{equation}
where $\{\beta_k \in (0,1)\}_{k=1}^K$ is a pre-defined variance schedule. Using the reparameterization trick, $\pmb{\mathcal{X}}_k$ can be sampled directly from $\pmb{\mathcal{X}}_0$~\cite{11278649} as
\begin{equation}
    \pmb{\mathcal{X}}_k
    =
    \sqrt{\bar{\alpha}_k}\,\pmb{\mathcal{X}}_0
    +
    \sqrt{1-\bar{\alpha}_k}\,\boldsymbol{\epsilon},
    \qquad
    \boldsymbol{\epsilon}\sim\mathcal{N}(\mathbf{0},\mathbf{I}),
\end{equation}
where $\bar{\alpha}_k=\prod_{s=1}^k \alpha_s$ and $\alpha_k=1-\beta_k$. As $K\to\infty$, $\pmb{\mathcal{X}}_K$ approaches an isotropic Gaussian distribution.

\subsubsection{Physics-Guided Reverse Process}
The reverse process learns the conditional transition distribution, $p_\theta(\pmb{\mathcal{X}}_{k-1}|\pmb{\mathcal{X}}_k,\mathbf{C}_n)$, where $\mathbf{C}_n$ denotes the conditioning tensor for slot $n$. Consistent with our early-fusion architecture, $\mathbf{C}_n$ is constructed by concatenating the delayed sparse observation, the corresponding sampling mask, and the current physics tensor, which can be expressed as
\begin{equation}
    \mathbf{C}_n
    =
    \mathcal{C}\!\left(
    \mathcal{Z}(\pmb{\mathcal{Y}}_{n-d}),
    \pmb{\mathcal{M}}_{n-d},
    \pmb{\mathcal{P}}_n
    \right),
\end{equation}
where $\mathcal{C}(\cdot)$ and $\mathcal{Z}(\cdot)$ denote the concatenation and zero-filling operations, respectively. Including binary mask $\pmb{\mathcal{M}}_{n-d}$ allows the neural network to distinguish unobserved locations from actually zero-valued signals.

Different from standard diffusion models that predict the added noise, our network $\hat{\pmb{\mathcal{X}}}_\theta$ is parameterized to directly reconstruct clean channel tensor $\pmb{\mathcal{X}}_0$ from noisy input $\pmb{\mathcal{X}}_k$ and conditioning tensor $\mathbf{C}_n$. The reverse transition is modeled as
\begin{equation}
    p_\theta(\pmb{\mathcal{X}}_{k-1}\mid \pmb{\mathcal{X}}_k,\mathbf{C}_n)
    =
    \mathcal{N}\!\left(
    \pmb{\mathcal{X}}_{k-1};
    \boldsymbol{\mu}_\theta(\pmb{\mathcal{X}}_k,k,\mathbf{C}_n),
    \tilde{\beta}_k \mathbf{I}
    \right),
\end{equation}
where $\tilde{\beta}_k=\frac{1-\bar{\alpha}_{k-1}}{1-\bar{\alpha}_k}\beta_k$ is the posterior variance. The mean, $\boldsymbol{\mu}_\theta$, is estimated using the tractable posterior distribution conditioned on the network's prediction of clean data, given by:
\begin{equation}
    \boldsymbol{\mu}_\theta
    =
    \frac{\sqrt{\bar{\alpha}_{k-1}}\beta_k}{1-\bar{\alpha}_k}
    \hat{\pmb{\mathcal{X}}}_\theta(\pmb{\mathcal{X}}_k,k,\mathbf{C}_n)
    +
    \frac{\sqrt{\alpha_k}(1-\bar{\alpha}_{k-1})}{1-\bar{\alpha}_k}
    \pmb{\mathcal{X}}_k.
\end{equation}
In this way, the delayed pilot observation and the current physical prior jointly guide the denoising process to enable accurate reconstruction of the real-time CSI.

\subsection{Proposed Neural Network Architecture}
\begin{figure}[!t]
    \centering
    \includegraphics[width=\linewidth]{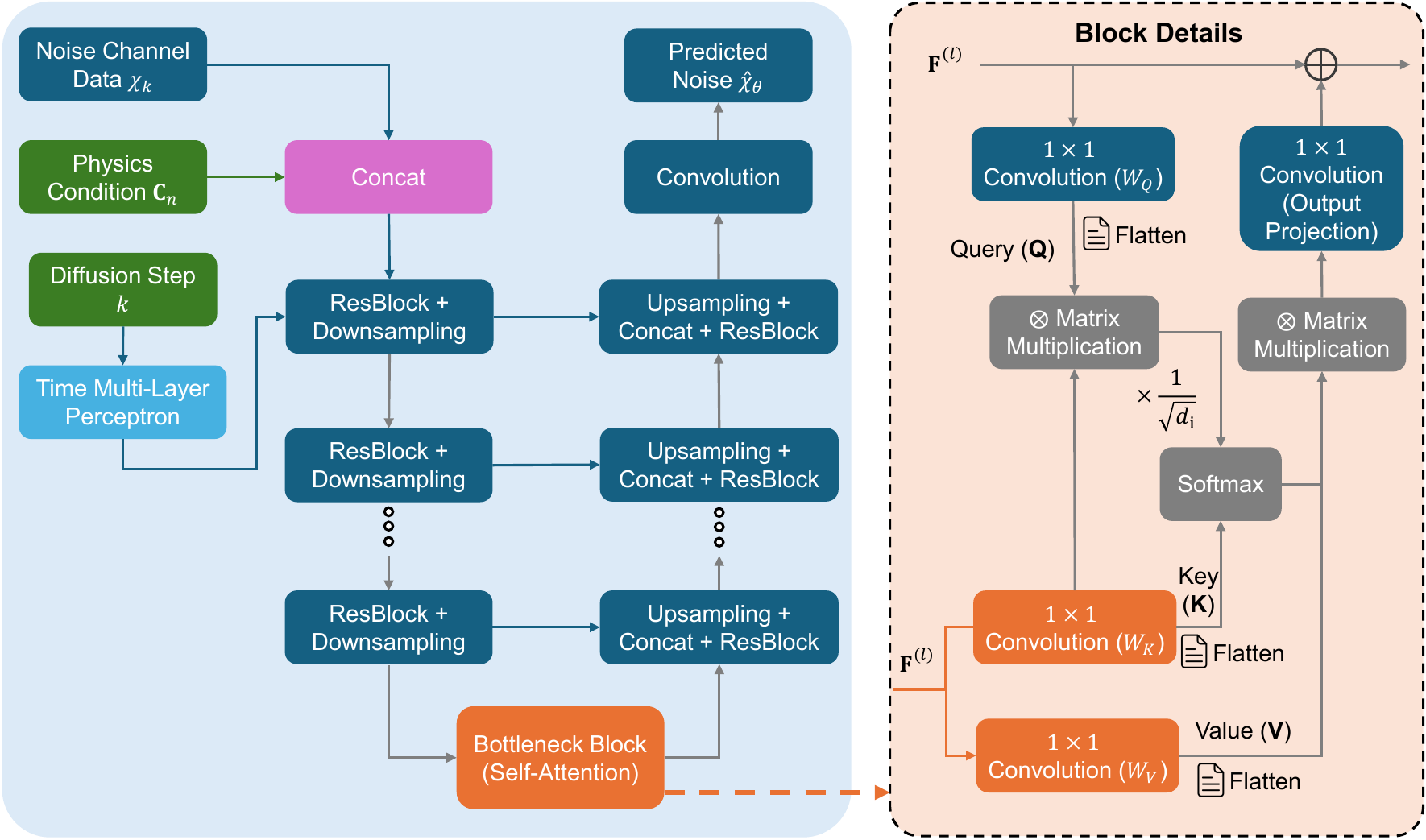}
    \caption{The proposed PG-ResUNet architecture.}
    \label{Unet}
\end{figure}

To effectively fuse the macroscopic physical constraints with the microscopic channel textures, we design a physics-guided residual U-Net (PG-ResUNet), as shown in Fig.~\ref{Unet}. Different from standard diffusion models that typically predict noise component $\boldsymbol{\epsilon}_\theta$, our network is parameterized to directly reconstruct clean channel state $\hat{\pmb{\mathcal{X}}}_\theta$. It takes noisy channel state $\pmb{\mathcal{X}}_k$, diffusion step $k$, and physical condition $\mathbf{C}_n$ as inputs to iteratively refine the signal.

\subsubsection{Physics-Aware Backbone with Early Fusion}
Our architecture adopts an early fusion strategy to deeply integrate physical constraints. Specifically, physical condition tensor $\mathbf{C}_n$ is concatenated with noisy channel data $\pmb{\mathcal{X}}_k$ along the channel dimension at the network input. This combined input is then processed by the encoder. We use a ResNet-based backbone to prevent gradient vanishing and ensure robust feature extraction. To integrate the time-dependent nature of the diffusion process, we utilize sinusoidal position embeddings to encode discrete time step $k$ into a continuous vector. This embedding is injected into each residual block (ResBlock) via a dense layer to modulate the feature response. Each ResBlock consists of group normalization, SiLU activation, and convolutional layers with residual connections to ensure stable training even with limited data.

\subsubsection{Bottleneck Self-Attention}
To capture long-range dependencies and complex non-linear interactions within the compressed feature space, we embed a self-attention module at the bottleneck layer of the U-Net. Let $\mathbf{F}^{(l)} \in \mathbb{R}^{H_l \times W_l \times C_l}$ denote the intermediate feature map at bottleneck layer $l$. This feature map encodes both the noisy signal structure and the physical priors through early fusion and convolutional operations. The interaction is modeled using a standard self-attention mechanism, where query $\mathbf{Q}^{(l)}$, key $\mathbf{K}^{(l)}$, and value $\mathbf{V}^{(l)}$ are all derived from the same feature map $\mathbf{F}^{(l)}$. 

We first flatten the spatial dimensions of the feature map into a sequence, $\tilde{\mathbf{F}}^{(l)} \in \mathbb{R}^{L_l \times C_l}$, where $L_l = H_l \times W_l$. The attention components are computed via linear projections, given by
\begin{equation}
    \mathbf{Q}^{(l)} = \tilde{\mathbf{F}}^{(l)} \mathbf{W}_Q^{(l)}, \mathbf{K}^{(l)} = \tilde{\mathbf{F}}^{(l)} \mathbf{W}_K^{(l)}, \mathbf{V}^{(l)} = \tilde{\mathbf{F}}^{(l)} \mathbf{W}_V^{(l)},
\end{equation}
where $\mathbf{W}_Q^{(l)}, \mathbf{W}_K^{(l)}, \mathbf{W}_V^{(l)} \in \mathbb{R}^{C_l \times d_\mathrm{i}}$ are learnable weight matrices. The attention output is then expressed as
\begin{equation}
    \mathcal{A}(\mathbf{Q}^{(l)}, \mathbf{K}^{(l)}, \mathbf{V}^{(l)}) = \mathcal{S}\left( \frac{\mathbf{Q}^{(l)} (\mathbf{K}^{(l)})^\mathrm{T}}{\sqrt{d_\mathrm{i}}} \right) \mathbf{V}^{(l)},
\end{equation}
where $\mathcal{S}(\cdot)$ is the softmax function. The output is projected back to original channel dimension $C_l$, reshaped, and passed through a residual connection. This mechanism allows the PG-ResUNet to dynamically weigh the importance of different spatial regions based on the fused physical-channel features.

\subsubsection{Objective Function}
The network is trained to minimize the discrepancy between ground-truth channel $\pmb{\mathcal{X}}_0$ and network output $\hat{\pmb{\mathcal{X}}}_{\theta}$. We formulate a hybrid loss function combining a weighted mean square error (MSE) and an L1 loss to ensure both pixel-level accuracy and structural fidelity, given by
\begin{align}
    \mathcal{L} &= \mathbb{E}_{\pmb{\mathcal{X}}_0, \mathbf{C}_n, \boldsymbol{\epsilon}, k} \left[ \left\| \mathbf{M}^{1/2} \odot (\pmb{\mathcal{X}}_0 - \hat{\pmb{\mathcal{X}}}_{\theta}(\pmb{\mathcal{X}}_k, k, \mathbf{C}_n)) \right\|_F^2 \right. \notag \\
    &\quad \left. + \lambda \| \pmb{\mathcal{X}}_0 - \hat{\pmb{\mathcal{X}}}_{\theta}(\pmb{\mathcal{X}}_k, k, \mathbf{C}_n) \|_1 \right],
\end{align}
where $\lambda$ is a hyperparameter balancing the L1 regularization term. To address the inherent sparsity of channel data, we introduce an amplitude-aware weighting tensor $\mathbf{M}$. Unlike binary sampling mask $\pmb{\mathcal{M}}$ used in the input, $\mathbf{M}$ is dynamically derived from the signal amplitude, which assigns higher weights to non-zero signal regions to force the model to prioritize the reconstruction of dominant channel paths.

\begin{figure}[!t]
    \centering
\includegraphics[width=\linewidth]{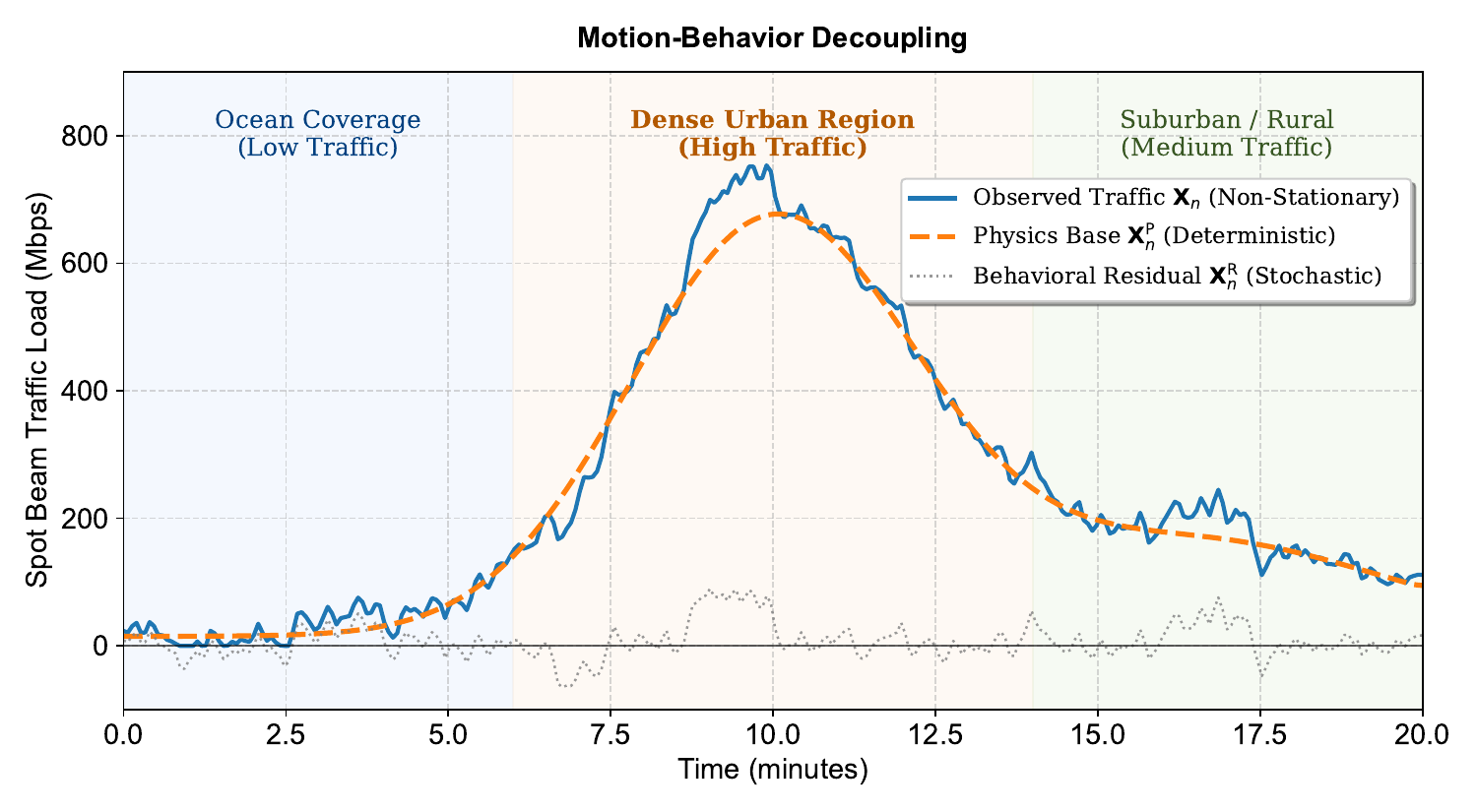}
    \caption{Illustration of the motion-behavior decoupling. The traffic trace is synthesized using the Hypatia simulator~\cite{hypatia}. The physics base is simulated based on the global population density data~\cite{worldpop2025}. The behavioral residual is simulated based on peak-hour internet traffic~\cite{sandvine2023global}.}
    \label{motion}
\end{figure}

\section{Traffic Digital Twin Construction}\label{tdt_sec}

Constructing a high-fidelity traffic DT is challenging due to mobility-induced non-stationarity. Fig.~\ref{motion} shows a representative traffic trace, where the satellite moves from an ocean region to a dense urban region. As a result, the observed traffic load over the spot beam changes sharply, rising from a low baseline (below 20~Mbps) to a high level (around 800~Mbps). This variation can be decomposed into two components: a deterministic motion-induced component, denoted by $\mathbf{X}_n^{\mathrm{P}}$, which is mainly governed by the satellite orbital motion and terrestrial population distribution, and a stochastic residual component, denoted by $\mathbf{X}_n^{\mathrm{R}}$, which captures random user behaviors. Conventional time-series models (e.g., standard LSTM) often perform poorly in this context, because they tend to misinterpret motion-driven topology changes as intrinsic traffic variations.

\begin{figure}[!t]
    \centering
\includegraphics[width=\linewidth]{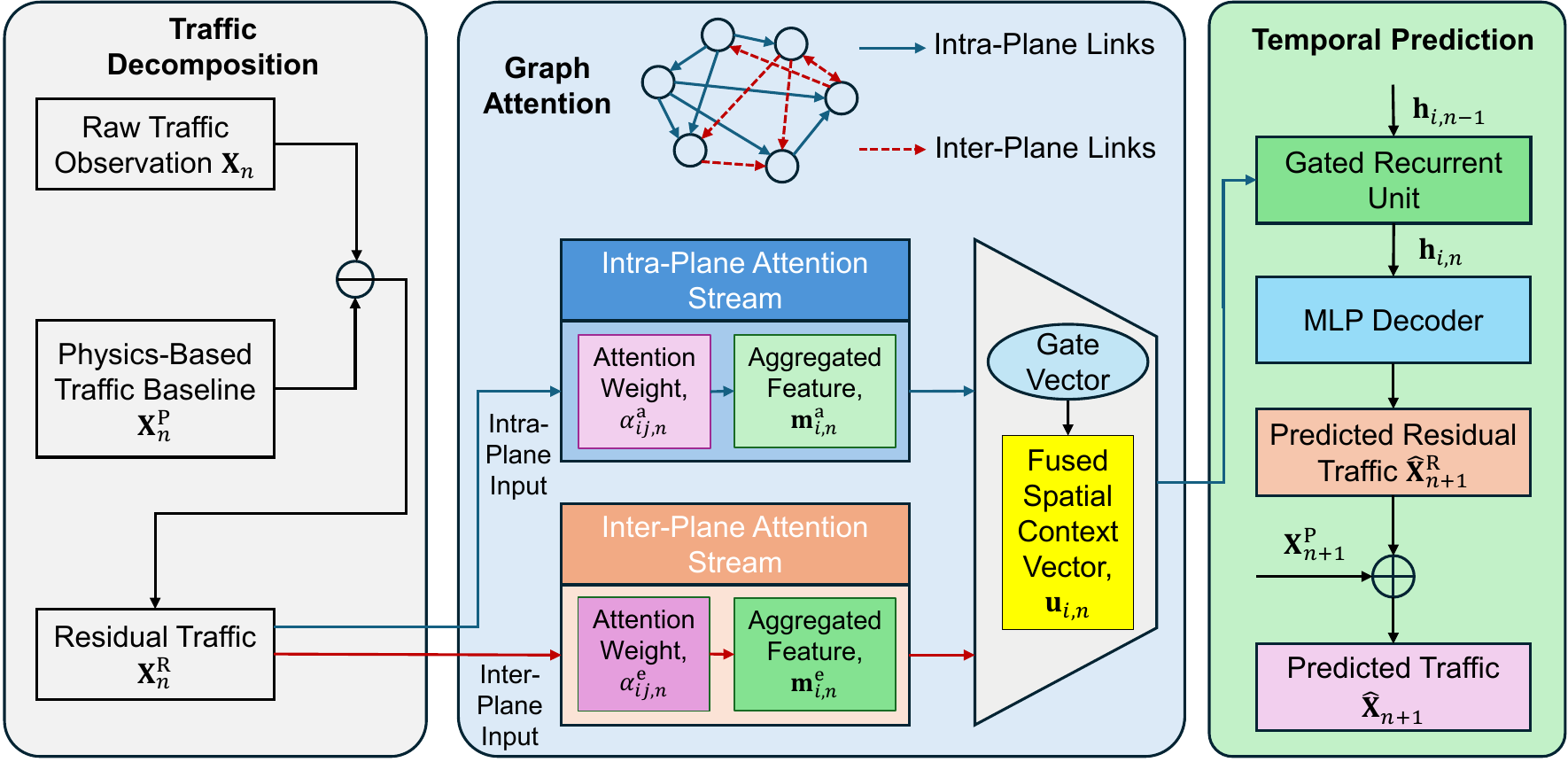}
    \caption{The proposed physics-informed graph neural network engine for the traffic DT. }
    \label{tdt}
\end{figure}

To address this challenge, we propose a physics-informed GNN for traffic DT construction, as shown in Fig.~\ref{tdt}. The proposed architecture consists of three modules: 1) traffic decomposition, which extracts stochastic residual $\mathbf{X}_n^{\mathrm{R}}$ by removing deterministic physics-based component $\mathbf{X}_n^{\mathrm{P}}$ from the observed traffic; 2) orbit-adaptive graph attention, which captures anisotropic spatial context through a dual-stream mechanism that distinguishes intra- and inter-plane dependencies; and 3) temporal prediction, which uses a gated recurrent unit (GRU) to model residual evolution and predicts next-slot traffic $\hat{\mathbf{X}}_{n+1}$ by combining the predicted residual with the corresponding physics-based component.

Let the LEO satellite constellation at slot $n$ be modeled as a dynamic graph $\mathcal{G}_n = (\mathcal{V}, \mathcal{E}_n, \mathbf{A}_n)$, where $\mathcal{V} = \{v_1, \dots, v_{N_{\mathrm{s}}}\}$ is the set of $N_{\mathrm{s}}$ satellites, $\mathcal{E}_n$ is the set of ISLs, and $\mathbf{A}_n \in \{0,1\}^{N_{\mathrm{s}} \times N_{\mathrm{s}}}$ is the adjacency matrix at slot $n$. The traffic load snapshot is denoted by $\mathbf{X}_n \in \mathbb{R}^{N_{\mathrm{s}} \times N_\text{b}}$, where $N_\text{b}$ is the number of spot beams served by each satellite. To mitigate mobility-induced non-stationarity, we decompose the observed traffic as
\begin{equation}
\mathbf{X}_n = \mathbf{X}_n^{\mathrm{P}}(\mathbf{s}_n,\mathbf{\Omega}) + \mathbf{X}_n^{\mathrm{R}},
\end{equation}
where $\mathbf{X}_n^{\mathrm{P}}$ is determined by satellite orbital state $\mathbf{s}_n$ and terrestrial population density distribution $\mathbf{\Omega}$, while $\mathbf{X}_n^{\mathrm{R}}$ captures the stochastic residual traffic. The prediction objective is to learn a mapping function $f_\varphi(\cdot)$, given by
\begin{equation}
\hat{\mathbf{X}}_{n+1}^{\mathrm{R}}
=
f_\varphi\!\left(
\mathbf{X}_{n-L+1:n}^{\mathrm{R}},
\mathcal{G}_{n-L+1:n}
\right),
\end{equation}
where $L$ is the look-back window length. The complete next-slot traffic is reconstructed as
\begin{equation}
\hat{\mathbf{X}}_{n+1}
=
\hat{\mathbf{X}}_{n+1}^{\mathrm{R}} + \mathbf{X}_{n+1}^{\mathrm{P}}.
\end{equation}

\subsection{Physics-Based Deterministic Modeling}

The physics-based deterministic component, $\mathbf{X}_n^{\mathrm{P}}$, serves as a dynamic baseline that captures the traffic demand induced purely by the satellite coverage geometry and the terrestrial population distribution. Specifically, let $\mathcal{B}_{m,b}(t_n) \subset \mathbb{R}^2$ denote the geographic footprint of the $b$-th spot beam of satellite $m$ at time $t_n$, where $m \in \{1,\dots,N_{\mathrm{s}}\}$ and $b \in \{1,\dots,N_{\mathrm{b}}\}$. This footprint is determined by the satellite orbital state and beamforming configuration at slot $n$.

We use a high-resolution global population density map, $\mathbf{\Omega} \in \mathbb{R}^{W \times H}$, with $\Omega(\mathbf{p})$ denoting the population density at terrestrial location $\mathbf{p}$. The deterministic traffic baseline for beam $b$ of satellite $m$ is then modeled as
\begin{equation}
    [\mathbf{X}_n^{\mathrm{P}}]_{m,b}
    =
    \rho
    \iint_{\mathbf{p} \in \mathcal{B}_{m,b}(t_n)}
    \Omega(\mathbf{p})
    \,\mathbb{I}\!\left(\epsilon_{m,b}(\mathbf{p}, t_n) \ge \epsilon_{\mathrm{th}}\right)
    \,\mathrm{d}\mathbf{p},
\end{equation}
where $\rho$ is a scaling factor representing the average baseline traffic demand per person, $\epsilon_{m,b}(\mathbf{p}, t_n)$ is the elevation angle from terrestrial location $\mathbf{p}$ to beam $b$ of satellite $m$ at time $t_n$, and $\mathbb{I}(\cdot)$ is the indicator function that retains only locations satisfying the minimum elevation requirements.

By evaluating this equation for all satellites and beams, we obtain baseline matrix $\mathbf{X}_n^{\mathrm{P}} \in \mathbb{R}^{N_{\mathrm{s}} \times N_{\mathrm{b}}}$. In this way, the static terrestrial demographic information is transformed into a dynamic satellite-centric traffic baseline that evolves with satellite orbital motion.

\subsection{Residual Learning via Orbit-Adaptive ST-GNN}

To capture stochastic user dynamics beyond the deterministic satellite orbital trend, we adopt a residual learning framework. By removing physics-based deterministic component $\mathbf{X}_n^{\mathrm{P}}$, the residual traffic matrix is given by
\begin{equation}
    \mathbf{X}_n^{\mathrm{R}} = \mathbf{X}_n - \mathbf{X}_n^{\mathrm{P}},
\end{equation}
where $\mathbf{X}_n^{\mathrm{R}}$ captures traffic fluctuations that cannot be explained by satellite orbital geometry alone. Based on a historical observation window, the goal is to predict next-slot residual traffic $\hat{\mathbf{X}}_{n+1}^{\mathrm{R}}$. To this end, we propose an orbit-adaptive ST-GNN, as shown in the right-hand modules of Fig.~\ref{tdt}. Unlike isotropic GNNs, the proposed orbit-adaptive ST-GNN distinguishes intra-plane and inter-plane dependencies to capture the structural anisotropy of LEO constellations.

Based on the previously defined dynamic graph $\mathcal{G}_n=(\mathcal{V},\mathcal{E}_n,\mathbf{A}_n)$, we partition the edge set into two disjoint subsets
\begin{equation}
    \mathcal{E}_n = \mathcal{E}_n^{\mathrm{a}} \cup \mathcal{E}_n^{\mathrm{e}},
\end{equation}
where $\mathcal{E}_n^{\mathrm{a}}$ denotes the set of intra-plane ISLs and $\mathcal{E}_n^{\mathrm{e}}$ denotes the set of inter-plane ISLs. The former mainly captures strong sequential dependencies induced by satellite orbital motion, while the latter provides additional spatial context under time-varying cross-plane connectivity.

Let $\mathbf{H}_n \in \mathbb{R}^{N_{\mathrm{s}} \times d_{\mathrm{h}}}$ denote the node feature matrix at slot $n$, where $d_{\mathrm{h}}$ is the hidden dimension. The hidden state of satellite $i$ is denoted by $\mathbf{h}_{i,n} \in \mathbb{R}^{d_{\mathrm{h}}}$. For the input layer, the node feature is initialized by the residual traffic vector of satellite $i$, denoted by $\mathbf{x}_{i,n}^{\mathrm{R}} \in \mathbb{R}^{N_{\mathrm{b}}}$.

To process anisotropic dependencies, we adopt a dual-stream attention mechanism. Let $\mathcal{N}_i^{\mathrm{a}}$ and $\mathcal{N}_i^{\mathrm{e}}$ denote the intra-plane and inter-plane neighbor sets of satellite $i$, respectively. For any neighbor $j$, the raw attention score is defined as
\begin{equation}
    e_{ij,n}
    =
    \mathcal{L}
    \!\left(
    \mathbf{a}^{\mathrm{T}}
    \left[
    \mathbf{W}\mathbf{h}_{i,n}
    \,\|\,
    \mathbf{W}\mathbf{h}_{j,n}
    \right]
    \right),
\end{equation}
where $\mathbf{W}\in\mathbb{R}^{d' \times d_{\mathrm{h}}}$ is a learnable weight matrix, $\mathbf{a}\in\mathbb{R}^{2d'}$ is a learnable attention vector, $\|$ denotes vector concatenation, and $\mathcal{L}(\cdot)$ is the leaky rectified linear unit activation, defined as $\mathcal{L}(x) = \max(0, x) + \gamma \min(0, x)$, where $\gamma$ is the negative slope parameter ensuring gradient flow for negative inputs.

To preserve the distinct statistical characteristics of different link types, we normalize the attention scores separately. Let $\phi \in \{\mathrm{a}, \mathrm{e}\}$ denote the link type. Then, the normalized attention weight is given by
\begin{equation}
    \alpha_{ij,n}^{\phi}
    =
    \frac{\exp(e_{ij,n})}
    {\sum\limits_{k \in \mathcal{N}_{i,n}^{\phi}} \exp(e_{ik,n})},
\end{equation}
where $\mathcal{N}_{i,n}^{\phi} = \{j \mid (i,j)\in\mathcal{E}_n^{\phi}\}$ denotes the neighbor set of satellite $i$ under link type $\phi$ at slot $n$. The aggregated neighborhood feature vector for link type $\phi$, denoted by $\mathbf{m}_{i,n}^{\phi}\in\mathbb{R}^{d'}$, is then given by
\begin{equation}
    \mathbf{m}_{i,n}^{\phi}
    =
    \sigma_{\mathrm{u}}
    \left(
    \sum_{j \in \mathcal{N}_i^{\phi}}
    \alpha_{ij,n}^{\phi}\mathbf{W}\mathbf{h}_{j,n}
    \right),
\end{equation}
where $\sigma_{\mathrm{u}}(\cdot)$ denotes the exponential linear unit activation. This yields two context vectors, namely $\mathbf{m}_{i,n}^{\mathrm{a}}$ and $\mathbf{m}_{i,n}^{\mathrm{e}}$, corresponding to intra-plane and inter-plane dependencies, respectively.

To prevent the denser intra-plane links from dominating the fused representation, we introduce a learnable gating mechanism. The gate vector, $\mathbf{g}_{i,n}\in\mathbb{R}^{d'}$, is defined as
\begin{equation}
    \mathbf{g}_{i,n}
    =
    \sigma_{\mathrm{o}}
    \left(
    \mathbf{W}_{\mathrm{g}}
    \left[
    \mathbf{m}_{i,n}^{\mathrm{a}}
    \,\|\,
    \mathbf{m}_{i,n}^{\mathrm{e}}
    \right]
    +
    \mathbf{b}_{\mathrm{g}}
    \right),
\end{equation}
where $\mathbf{W}_{\mathrm{g}}\in\mathbb{R}^{d' \times 2d'}$ and $\mathbf{b}_{\mathrm{g}}\in\mathbb{R}^{d'}$ are learnable parameters, and $\sigma_{\mathrm{o}}(\cdot)$ is the sigmoid activation function. The final fused spatial context vector, $\mathbf{u}_{i,n}\in\mathbb{R}^{d'}$, is given by
\begin{equation}
    \mathbf{u}_{i,n}
    =
    \mathbf{g}_{i,n}\odot \mathbf{m}_{i,n}^{\mathrm{a}}
    +
    (\mathbf{1}-\mathbf{g}_{i,n})\odot \mathbf{m}_{i,n}^{\mathrm{e}}.
\end{equation}

To model the temporal evolution of traffic residuals, we integrate spatial context $\mathbf{u}_{i,n}$ into a GRU. Update gate $\mathbf{z}_{i,n}\in\mathbb{R}^{d_{\mathrm{h}}}$ and reset gate $\mathbf{r}_{i,n}\in\mathbb{R}^{d_{\mathrm{h}}}$ are given by
\begin{align}
    \mathbf{z}_{i,n}
    &= \sigma_{\mathrm{o}}\!\left(
    \mathbf{W}_{\mathrm{z}}
    \left[
    \mathbf{x}_{i,n}^{\mathrm{R}}
    \,\|\,
    \mathbf{u}_{i,n}
    \right]
    +
    \mathbf{U}_{\mathrm{z}}\mathbf{h}_{i,n-1}
    +
    \mathbf{b}_{\mathrm{z}}
    \right), \\
    \mathbf{r}_{i,n}
    &= \sigma_{\mathrm{o}}\!\left(
    \mathbf{W}_{\mathrm{r}}
    \left[
    \mathbf{x}_{i,n}^{\mathrm{R}}
    \,\|\,
    \mathbf{u}_{i,n}
    \right]
    +
    \mathbf{U}_{\mathrm{r}}\mathbf{h}_{i,n-1}
    +
    \mathbf{b}_{\mathrm{r}}
    \right),
\end{align}
where $\mathbf{W}_\text{z}, \mathbf{W}_\text{r} \in \mathbb{R}^{d_\text{h} \times (N_\text{b} + d')}$ are learnable weight matrices that project the concatenated input into the hidden state space, $\mathbf{U}_\text{z}, \mathbf{U}_\text{r} \in \mathbb{R}^{d_\text{h} \times d_\text{h}}$ are recurrent weight matrices applied to previous hidden state $\mathbf{h}_{i,t-1}$, and $\mathbf{b}_\text{z}, \mathbf{b}_\text{r} \in \mathbb{R}^{d_\text{h}}$ denote bias vectors.

The hidden state update involves a two-step process. First, we analyze candidate hidden state $\tilde{\mathbf{h}}_{i,t}\in\mathbb{R}^{d_\text{h}}$, which represents the new information derived from the current input and the reset memory, i.e., 
\begin{equation}
    \tilde{\mathbf{h}}_{i,n} = \sigma_\text{h}\left( \mathbf{W}_\text{h} [\mathbf{x}_{i,n}^{\mathrm{R}} \, \| \, \mathbf{u}_{i,n}] + \mathbf{U}_\text{h} (\mathbf{r}_{i,n} \odot \mathbf{h}_{i,n-1}) + \mathbf{b}_\text{h} \right),
\end{equation}
where $\mathbf{W}_\text{h} \in \mathbb{R}^{d_\text{h} \times (N_\text{b} + d')}$ is the weight matrix mapping the concatenated input to the hidden space, $\mathbf{U}_\text{h} \in \mathbb{R}^{d_\text{h} \times d_\text{h}}$ is the recurrent weight matrix governing the influence of the previous hidden state (modulated by the reset gate) on the current candidate state, and $\mathbf{b}_\text{h} \in \mathbb{R}^{d_\text{h}}$ is the bias vector. Function $\sigma_\text{h}(\cdot)$ denotes the hyperbolic tangent activation function ($\tanh$) to ensure the state values remain within $[-1, 1]$.

Then, the final hidden state, $\mathbf{h}_{i,n}\in\mathbb{R}^{d_\text{h}}$, is obtained by linearly interpolating between the previous state and the candidate state via the update gate, given by
\begin{equation}
    \mathbf{h}_{i,n} = (\mathbf{1} - \mathbf{z}_{i,n}) \odot \mathbf{h}_{i,n-1} + \mathbf{z}_{i,n} \odot \tilde{\mathbf{h}}_{i,n}.
\end{equation}
This mechanism allows the model to adaptively capture long-term dependencies by preserving past information when the elements of $\mathbf{z}_{i,n}$ are small, or capturing sudden traffic bursts when they approach $1$.

\section{Experiment Results}\label{exp}
In this section, we present a comprehensive evaluation of the proposed physics-informed NTN DT framework.\footnote{The experimental code and datasets will be released after the paper is published.}
\subsection{Simulation Setup}

\subsubsection{LEO Constellation Construction}
\begin{figure}[!t]
    \centering
\includegraphics[width=\linewidth]{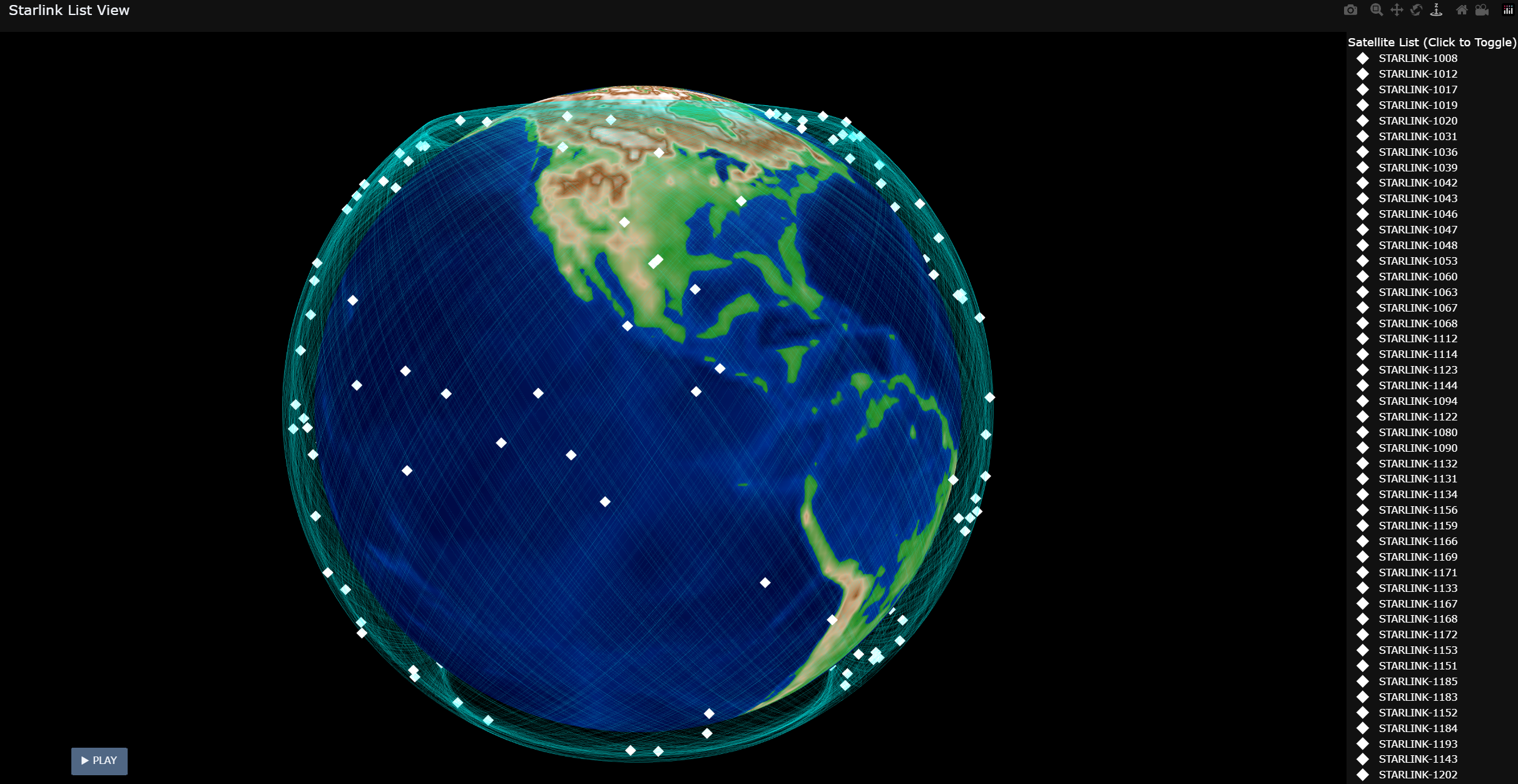}
    \caption{The constructed LEO constellation based on real-world Starlink data (Sampling 200 satellites for view).}
    \label{orbit}
\end{figure}
To establish a high-fidelity dynamic network topology, we construct the LEO constellation simulation based on real-world orbital parameters from the Starlink mega-constellation, as shown in Fig.~\ref{orbit}. The construction process involves three key stages: data acquisition, orbital propagation, and coordinate transformation. First, the initial orbital states are derived from the latest two-line element (TLE) sets retrieved from the CelesTrak database.\footnote{\url{https://celestrak.org/NORAD/elements/gp.php?GROUP=starlink&FORMAT=tle}} A subset of $N_\text{s} = 1,000$ satellites is selected to represent a dense orbital shell. Second, to capture the temporal evolution of the network, we use the SGP4 propagator implemented via the Skyfield astrodynamics library. The SGP4 model is chosen for its ability to accurately predict satellite positions by accounting for secular and periodic variations caused by Earth's oblateness and atmospheric drag. The simulation spans a total duration of $8$ hours with a fine-grained sampling interval of $\Delta t = 0.1$ seconds to generate high-resolution trajectory data. For each time step $t$, the satellite's position and velocity vectors are computed in the geocentric inertial frame and subsequently transformed into the ECEF reference frame. %This transformation ensures spatial alignment with the terrestrial user grid for subsequent channel modeling. 
The final trajectory tensor $\mathcal{T} \in \mathbb{R}^{N_\text{s} \times T \times 6}$ contains the instantaneous position $\mathbf{r}(t)$ and velocity $\mathbf{v}(t)$ for all nodes.

\subsubsection{Physics-Aware LEO Channel Digital Twin Construction}
To build a high-fidelity simulation environment, we construct a physics-aware LEO channel DT that synergies dynamic orbital kinematics with real-world atmospheric impairments. The simulation is grounded in a precise link budget calculation driven by the Starlink constellation's ephemeris, where satellite trajectories are propagated using the SGP4 model to derive instantaneous slant ranges and elevation angles for baseline FSPL and Doppler shift estimation at the Ku-band (12 GHz). To transcend idealized statistical models, we integrate site-specific weather dynamics from the ECMWF ERA5 reanalysis dataset.\footnote{\url{https://cds.climate.copernicus.eu/datasets/reanalysis-era5-single-levels?tab=download}} We map precipitation rates to rain attenuation via the ITU-R P.838-3 model~\cite{838} and synthesize tropospheric scintillation using the ITU-R P.618 recommendation~\cite{618}. %The latter combines a wet refractive component derived from thermodynamic variables with a dry turbulence floor to capture fast-fading stochasticity.
The final CSI is synthesized by superimposing these deterministic geometric losses with stochastic atmospheric attenuation to capture both macroscopic satellite orbital trends and microscopic weather-induced fluctuations.

Building upon this high-fidelity simulation environment, we implement the proposed PCDM as the core generative engine. %The PCDM architecture is founded on a PG-ResUNet backbone to learn the conditional distribution $p_\theta(\mathcal{X}_0 | \mathbf{C}_t)$. 
To ensure physical consistency, the model is conditioned on a composite tensor $\mathbf{C}_n$, which concatenates historical CSI observations with environmental priors, such as normalized path loss, rain attenuation maps, and Doppler shifts. Temporal dynamics are encoded via sinusoidal time embeddings injected into each residual block, while a physics-aware attention mechanism captures long-range spatial dependencies within the grid. The training process uses a 1000-step denoising diffusion probabilistic model optimized via the AdamW optimizer with cosine annealing scheduling, to ensure stable convergence in the high-dimensional channel state space. The detailed hyperparameter configuration is provided in Table~\ref{tab:pcdm_config}.
\begin{table}[!t]
\centering
\caption{Detailed PCDM Hyperparameter Configuration}
\label{tab:pcdm_config}
\begin{tabular}{ll}
\toprule
\textbf{Parameter} & \textbf{Configuration} \\
\midrule
\multicolumn{2}{l}{\textit{A. PG-ResUNet Architecture}} \\
\midrule
Backbone Structure & Symmetric Encoder-Decoder \\
Base Channel Dimension & 64 \\
Channel Multipliers & 64 $\to$ 128 $\to$ 256 \\
Encoder/Decoder Levels & 3 Levels \\
Residual Block Type & GroupNorm (8) + SiLU + Conv 3x3 \\
Time Embedding Dimension & 256 (Sinusoidal) \\
Attention Mechanism & Self-Attention (Heads=4) at Bottleneck \\
\midrule
\multicolumn{2}{l}{\textit{B. Input Dimensions}} \\
\midrule
Grid Resolution & $64 \times 64$ \\
Target Input Channels & 2 (Real \& Imaginary CSI) \\
Condition Channels & 9 (History(2) + Physics(6) + Mask(1)) \\
Total Input Channels & 11 \\
\midrule
\multicolumn{2}{l}{\textit{C. Diffusion \& Training Hyperparameters}} \\
\midrule
Diffusion Steps  & 1000 \\
Noise Schedule & Linear ($\beta_{\min}=10^{-4}, \beta_{\max}=0.02$) \\
Loss Function & $\mathcal{L}_\text{MSE} + 0.05 \cdot \mathcal{L}_{L1}$ \\
Optimizer & AdamW \\
Learning Rate & $1 \times 10^{-4}$ \\
Learning Rate Scheduler & Cosine Annealing ($\eta_{\min}=10^{-6}$) \\
Batch Size & 256 \\
Total Epochs & 50 \\
Precision & Mixed Precision (BFloat16) \\
\bottomrule
\end{tabular}
\end{table}

\begin{figure}[!t]
    \centering
\includegraphics[width=0.8\mysinglefigwidth]{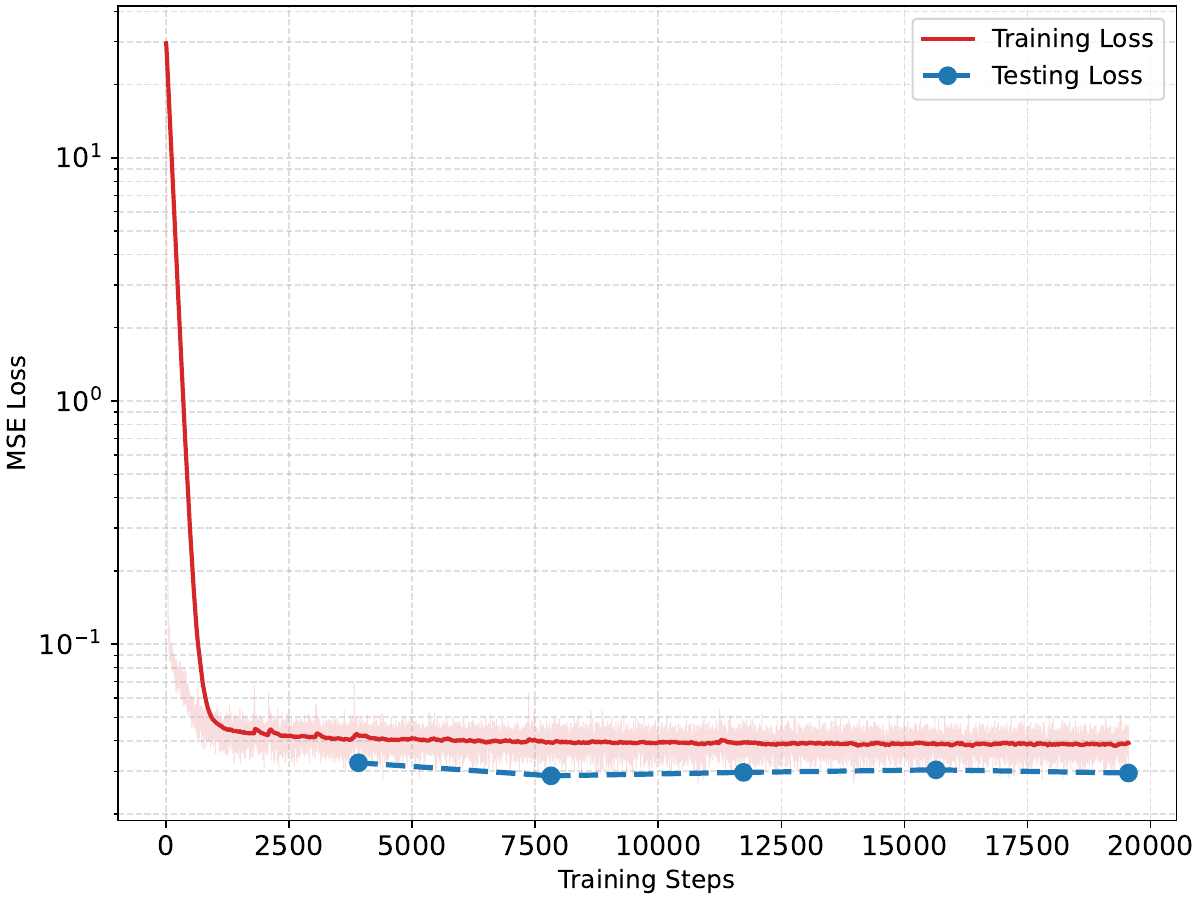}
    \caption{Training and testing loss of the proposed PCDM in the channel DT.}
    \label{pcdm_loss}
\end{figure}

To verify the effectiveness of the proposed physics-guided architecture, we monitored the training/testing stability and convergence behavior. As shown in Fig.~\ref{pcdm_loss}, the training loss exhibits a sharp descent within the initial 2,500 iterations, driven by the effectiveness of the AdamW optimizer and the inductive biases provided by the physical condition inputs. Despite the high stochasticity of the LEO channel data (e.g., intermittent rain events and rapid Doppler variations), the model achieves a stable asymptotic MSE of approximately $4 \times 10^{-2}$ without exhibiting divergence. The smooth decay of the loss curve guarantees that the PCDM effectively captures the underlying distribution of the physics-aware channel DT.

\begin{figure}[!t]
    \centering
\includegraphics[width=\mysinglefigwidth]{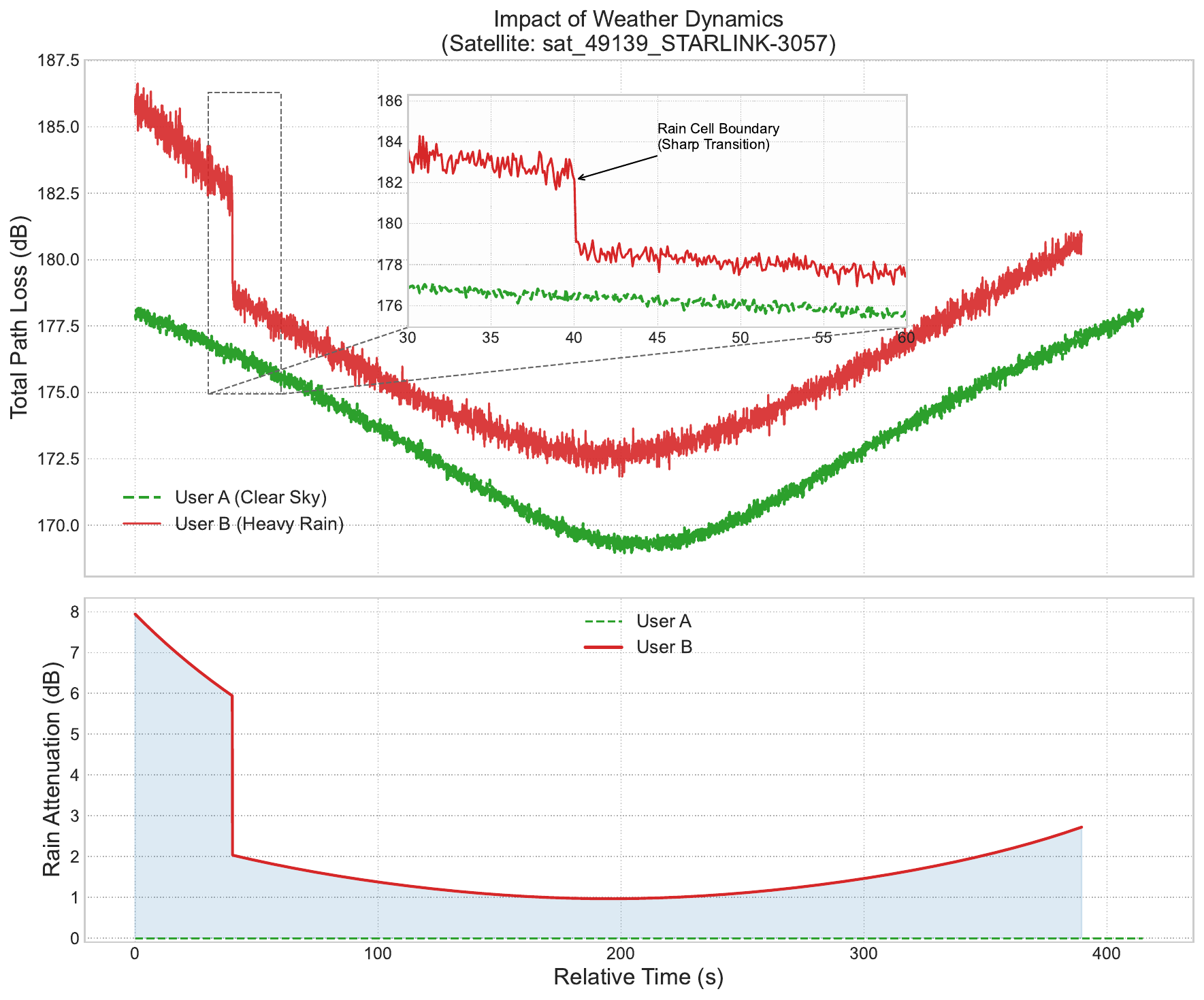}
    \caption{Time-varying channel dynamics visualization.}
    \label{channel}
\end{figure}

To validate the physical fidelity of our constructed channel DT, we visualize the temporal channel evolution during a typical Starlink satellite pass in Fig.~\ref{channel}. The simulation results highlight several distinct physical characteristics modeled by our engine. Specifically, both users exhibit a smooth U-shaped baseline trajectory. This corresponds to the dominance of FSPL governed by the satellite's kinematic elevation changes, which validates the accuracy of our SGP4-based orbital propagation. A sharp transition in path loss is observed for User B at $t \approx 40$ seconds. This step change is perfectly aligned with the jump in the rain attenuation profile, which captures the physical event of a link path crossing the boundary of a distinct rain cell.

\subsubsection{Physics-Aware LEO Traffic Digital Twin Construction}
To simulate realistic network loads under dynamic LEO coverage, we develop a high-fidelity traffic generation engine that couples satellite orbital kinematics with the 2025 global population distribution.\footnote{\url{https://hub.worldpop.org/geodata/summary?id=80031}} The process begins with geometric mapping, where satellite spot beam coverage is simulated via a spatial pooling mechanism that samples peak population densities within a $0.1^\circ$ window of the nadir point to capture maximum demand. To account for inter-network competition, we introduce an urban suppression factor that inversely scales satellite penetration rates based on population density (e.g., restricting urban usage to 5\% while assuming 80\% penetration in rural areas), which can effectively dampen traffic in core cities while amplifying rural demand. Finally, temporal dynamics are synthesized by convolving the physical baseline with a Hanning window to model beam integration effects, and superimposing a stochastic behavioral component driven by an AR-1 process to capture non-stationary burstiness with strong temporal memory.
%($n_t = 0.95 n_{t-1} + \epsilon_t$)
\begin{figure}[!t]
    \centering
\includegraphics[width=\mysinglefigwidth]{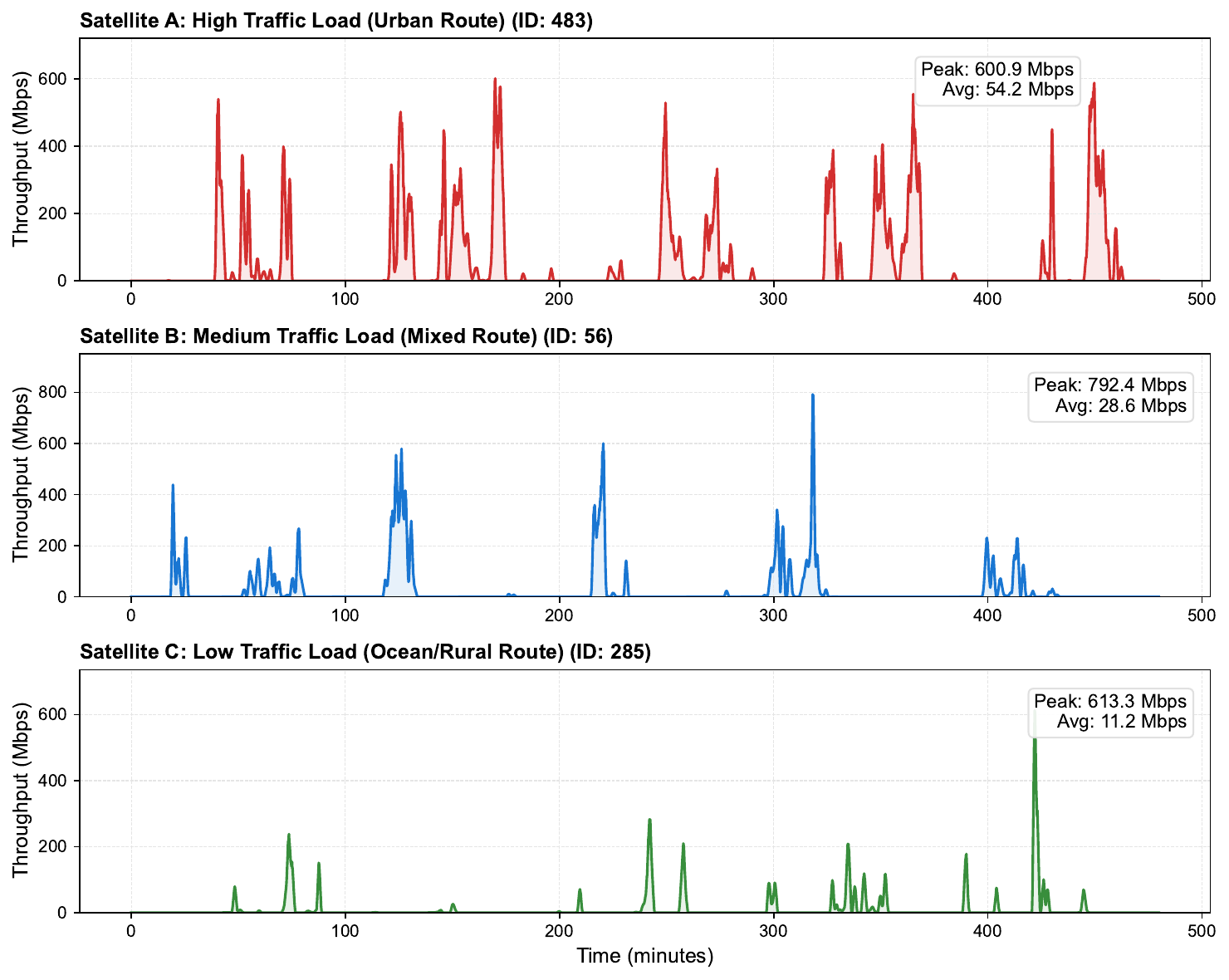}
    \caption{Spatiotemporal traffic diversity validation across different satellite orbits.}
    \label{traffic}
\end{figure}

To validate the heterogeneity of the generated traffic dataset, we visualize the temporal throughput evolution of three representative satellites in Fig.~\ref{traffic}. The simulation results demonstrate significant spatiotemporal diversity driven by the underlying demographic constraints. \textit{High-Load Urban Route:} Satellite A (ID: 483, top panel), passing over core urban agglomerations, exhibits frequent and high-amplitude bursts with peak rates exceeding 600 Mbps. This confirms that the spatial pooling mechanism effectively captures the aggregated demand of dense user clusters. \textit{Sparse Rural/Oceanic Route:} In contrast, Satellite C (ID: 285, bottom panel) traverses predominantly oceanic or remote rural regions. Its traffic profile is characterized by long periods of silence punctuated by sporadic low-intensity spikes, reflecting the urban suppression logic where service penetration is high but absolute user density is low. \textit{Non-Stationarity:} Across all scenarios, the traffic exhibits strong non-stationarity and varying peak-to-average ratios (e.g., Sat A: $\approx 11$, Sat C: $\approx 54$), which validates the requirement for the subsequent orbit-adaptive graph learning approach to capture these complex dynamics.

To accurately predict the stochastic behavioral residual component $\mathbf{X}_n^\mathrm{R}$ separated from the physical baseline, we deploy an orbit-adaptive ST-GNN in the traffic DT. Unlike static graph networks, our architecture is explicitly designed to handle the dynamic topology of LEO constellations where neighbor relationships evolve due to orbital kinematics. The model architecture integrates three critical mechanisms: (1) To capture spatial correlations among satellites, the model dynamically constructs an adjacency matrix $A_n$ at each time slot. Instead of using fixed Euclidean distances, we map the real-time satellite positions (Lat/Lon) into 3D spherical coordinates and apply a k-nearest neighbor (k-NN) algorithm ($k=4$) to identify the most relevant spatial neighbors; (2) The temporal dependencies are modeled by a stacked GRU (2 layers), which captures the time-varying traffic patterns. This is coupled with graph convolutional network (GCN) layers that aggregate features from the dynamic neighbors identified by the k-NN graph, followed by layer normalization and residual connections to prevent gradient vanishing; (3) The input tensor is augmented with explicit satellite orbital information. The latitude and longitude are encoded via trigonometric transformations ($\sin(\cdot), \cos(\cdot)$) to preserve the cyclic nature of the spherical geometry, which generates a 5-dimensional feature vector per node (1 traffic residual + 4 coordinate embeddings). Table \ref{tab:stgnn_config} summarizes the detailed hyperparameter settings.
\begin{table}[!t]
\centering
\caption{Hyperparameter configuration of the orbit-adaptive ST-GNN}
\label{tab:stgnn_config}
\begin{tabular}{ll}
\toprule
\textbf{Parameter} & \textbf{Configuration} \\
\midrule
\multicolumn{2}{l}{\textit{A. Neural Network Architecture}} \\
\midrule
Model Type & Hybrid GRU-GCN \\
Input Dimension & 5 \\
Hidden Dimension & 128 \\
Number of Layers & 2 (Stacked GRU + GCN) \\
Graph Construction & Dynamic k-NN on spherical coordinates \\
Output Head & MLP (Linear$\to$ReLU$\to$Dropout$\to$Linear) \\
Dropout Rate & 0.2 \\
\midrule
\multicolumn{2}{l}{\textit{B. Data \& Input Sequencing}} \\
\midrule
Look-back Window & 12 time steps \\
Normalization & Standard scaler (Zero mean, unit var) \\
Coordinate Encoding & Trigonometric transform ($\sin, \cos$) \\
\midrule
\multicolumn{2}{l}{\textit{C. Training Optimization}} \\
\midrule
Optimizer & Adam \\
Learning Rate & $2 \times 10^{-3}$ \\
Learning Rate Scheduler & Cosine annealing ($T_{\max}=250$) \\
Loss Function & MSE \\
Batch Size & 32 \\
Total Epochs & 250 \\
Precision & Mixed precision (AMP / GradScaler) \\
\bottomrule
\end{tabular}
\end{table}

\begin{figure}[!t]
    \centering
    \includegraphics[width=\linewidth]{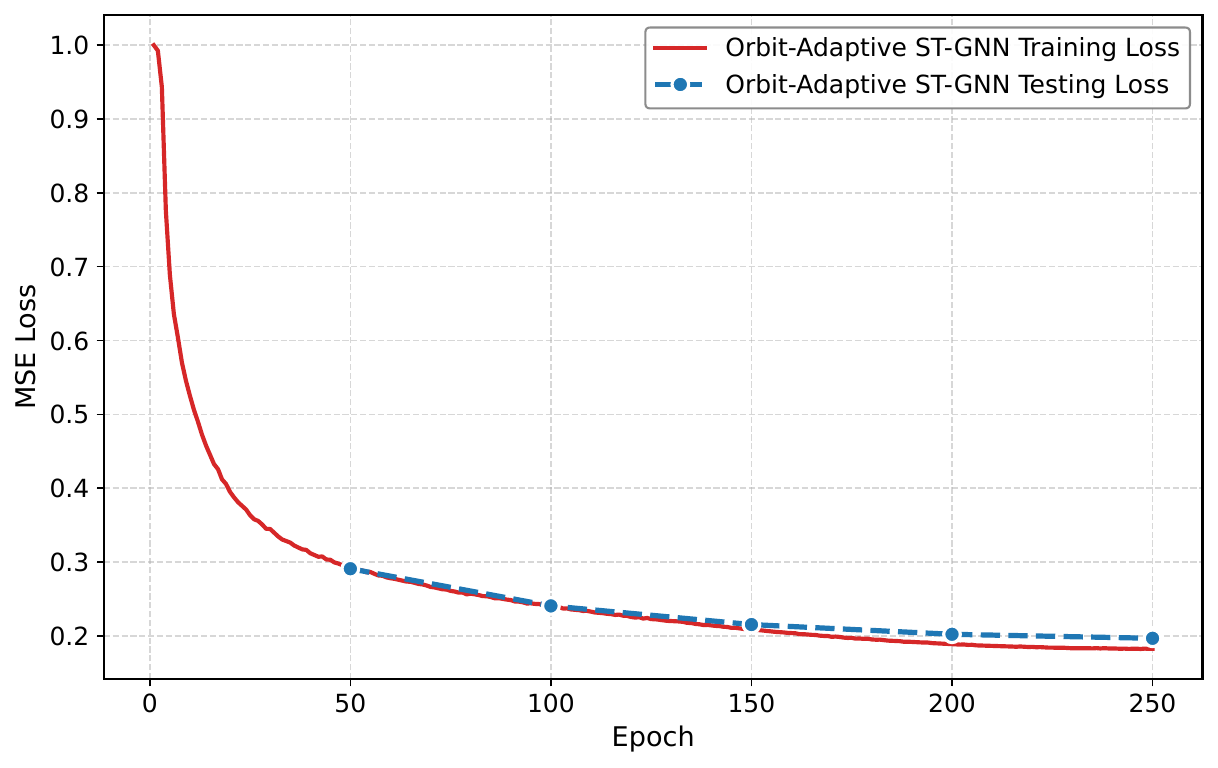}
    \caption{Training and testing convergence of the proposed orbit-adaptive ST-GNN in the traffic DT.}
    \label{fig:stgnn_loss}
\end{figure}

To validate the training/testing stability of the proposed graph learning framework, we track the MSE loss evolution over the entire optimization process. As depicted in Fig. \ref{fig:stgnn_loss}, the proposed orbit-adaptive ST-GNN exhibits a rapid convergence characteristic within the first 50 epochs, driven by the effective gradient propagation through the residual connections in the GRU-GCN layers. Under the cosine annealing learning rate scheduler, the loss curve smoothly transitions from the initial exploration phase to a stable fine-tuning stage, ultimately converging to an MSE of approximately 0.18 without signs of overfitting or oscillation. This confirms that the dynamic spherical graph construction successfully captures the underlying spatiotemporal correlations of non-stationary traffic data.

\begin{figure}[!t]
    \centering
    \includegraphics[width=\linewidth]{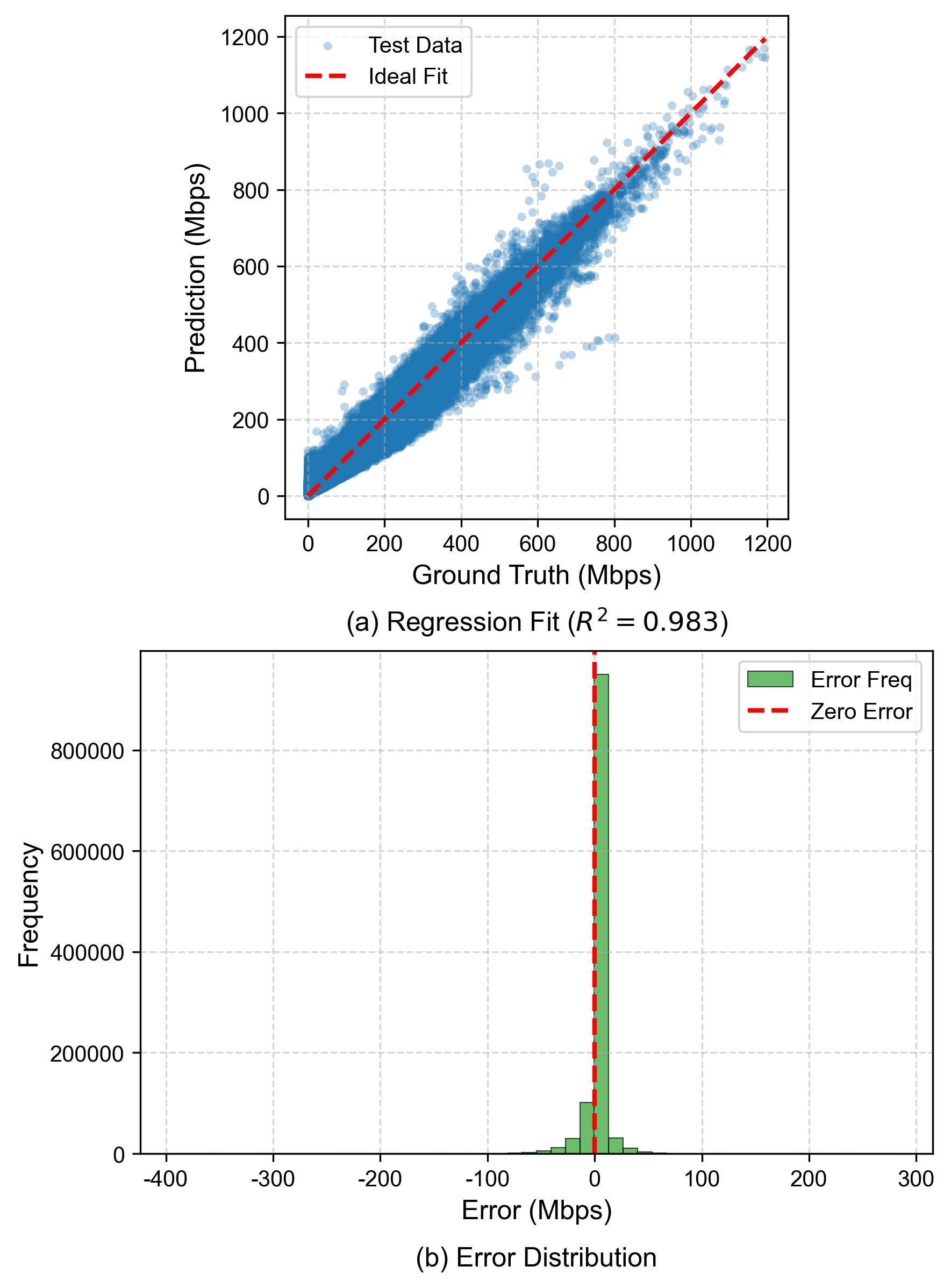}
    \caption{Statistical error analysis of the ST-GNN predictor in the traffic DT.}
    \label{fig:error_analysis}
\end{figure}

\begin{figure*}[!t]
    \centering
    \includegraphics[width=\linewidth]{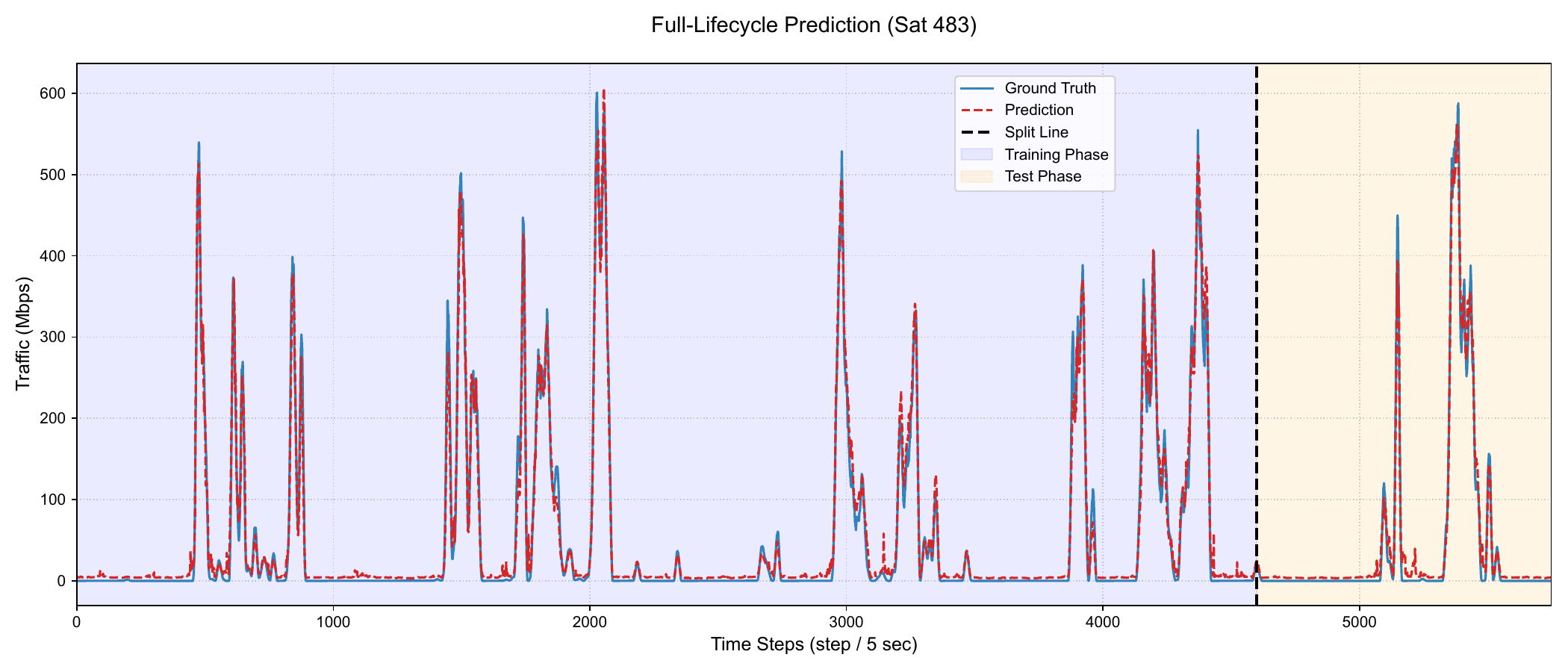}
    \caption{Full lifecycle traffic prediction visualization for a high-traffic satellite (Sat-483) in the traffic DT.}
    \label{fig:lifecycle}
\end{figure*}

To further quantify the statistical fidelity of the prediction engine in the traffic DT, we conduct a comprehensive error analysis. Fig. \ref{fig:error_analysis} presents the regression fit (left panel) and the error distribution histogram (right panel). The regression analysis reveals an exceptionally strong linear correlation between the predicted values and the ground truth, which achieves a coefficient of determination ($R^2$) of 0.983. Furthermore, the error distribution is symmetric and centered at zero, resembling a Gaussian distribution. This confirms that the proposed orbit-adaptive ST-GNN serves as an unbiased estimator to effectively minimize forecasting errors while handling the high-variance stochasticity of the satellite traffic.

To qualitatively evaluate the trained model's generalization capability, Fig. \ref{fig:lifecycle} shows the full-lifecycle prediction results for a representative high-load satellite (Sat-483). The timeline is vertically partitioned into a training phase (light blue background) and a testing phase (light yellow background). It is evident that the predicted traffic trajectory (red dashed line) maintains high fidelity to the ground truth (blue solid line) even in the unseen testing phase. Moreover, the proposed orbit-adaptive ST-GNN accurately captures both the sporadic silence intervals (traffic $\approx$ 0) and the extreme bursty peaks (traffic $>$ 500 Mbps), which demonstrates that the model has successfully learned the underlying spatiotemporal dependencies rather than merely memorizing historical patterns.

\subsection{Performance Comparison in the Channel Digital Twin}
To evaluate the generation quality and physical fidelity of the proposed PCDM, we benchmark it against two state-of-the-art deep generative models in the channel DT. 
\begin{itemize}
    \item Conditional GAN (cGAN)~\cite{banerjee2022downlink}: In the cGAN, the generator selects a physics-guided U-Net (PGUNet) architecture. Unlike standard U-Nets, it incorporates a physics-guided cross-attention module at the bottleneck ($8 \times 8$ resolution). This module computes attention scores between the channel features and the encoded physical conditions to explicitly guide the generation process. The discriminator adopts a PatchGAN architecture to enforce high-frequency texture consistency by penalizing local $N \times N$ patches. The neural network is trained using a composite objective combining adversarial MSE loss and a weighted L1 reconstruction loss.
    \item VAE-based linear MMSE (VAE-LMMSE)~\cite{kasibovic2025addressing}: It features a dual-head decoder that simultaneously predicts the channel mean and the diagonal covariance. The model utilizes a 128-dimensional latent space and is optimized via the evidence lower bound (ELBO), which comprises a Gaussian negative log-likelihood (NLL) term for reconstruction and a kullback-leibler (KL) divergence term ($\beta=0.001$) for regularization.
\end{itemize}

The specific architectural and training configurations for these baselines are detailed in Table~\ref{tab:baseline_config}.

\begin{table}[!t]
\centering
\caption{Baseline Configuration in the Channel DT}
\label{tab:baseline_config}
\begin{tabular}{l|cc}
\toprule
\textbf{Parameter} & \textbf{cGAN} & \textbf{VAE-LMMSE} \\
\midrule
\multicolumn{3}{l}{\textit{A. Architecture Specifications}} \\
\midrule
Core Network & PGUNet & Encoder-decoder \\
Feature Interaction & Cross-attention  & Concat \\
Latent/Noise Dim & Noise ($64\times64$) & Vector ($128$) \\
Discriminator & 3-layer Conv & N/A \\
\midrule
\multicolumn{3}{l}{\textit{B. Training Objectives}} \\
\midrule
Loss Function & $\mathcal{L}_\text{MSE}^\text{Adv} + 100 \cdot \mathcal{L}_{L1}$ & $\mathcal{L}_\text{NLL} + 0.001 \cdot \mathcal{L}_\text{KL}$ \\
Optimizer & Adam & AdamW \\
Learning Rate & \makecell{$G: 2\times10^{-4}$ \\ $D: 5\times10^{-5}$} & $1\times10^{-3}$ \\
Batch Size & 256 & 256 \\
Epochs & 50 & 100 \\
Precision & AMP & AMP \\
\bottomrule
\end{tabular}
\end{table}

\begin{figure*}[!t]
    \centering
    \includegraphics[width=\linewidth]{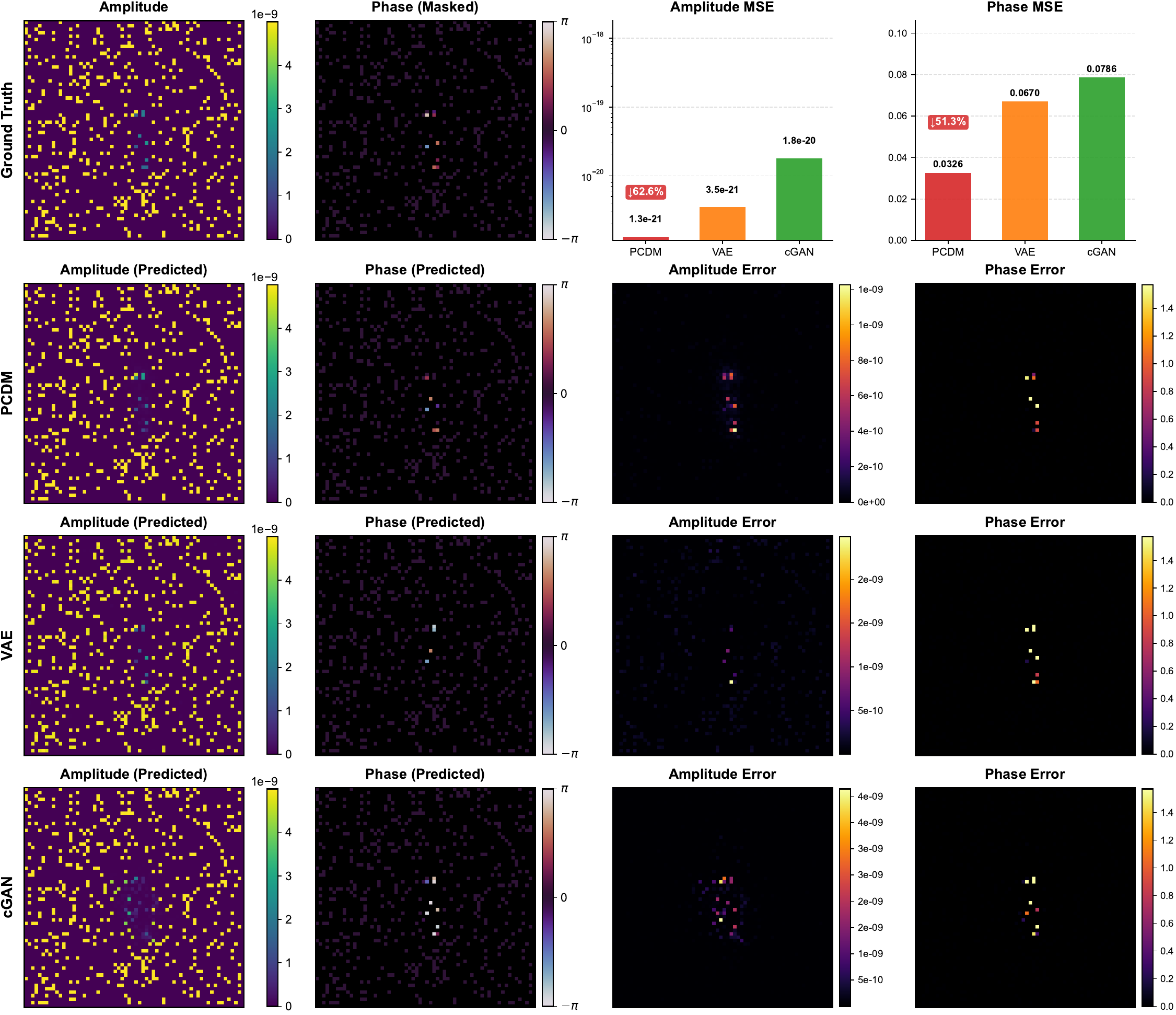}
    \caption{Visual and quantitative comparison of channel DT construction quality. Note that the ground-truth phase is masked by an amplitude threshold to filter out random-noise phases in zero-energy regions, thereby allowing clear visualization of the dominant multipath components.}
    \label{fig:channel_visual}
\end{figure*}

To intuitively assess the fidelity of the constructed channel DT, we present a comprehensive comparison of the generated CSI across different models in Fig.~\ref{fig:channel_visual}. The visualization is organized as follows: the top row shows the ground truth alongside the quantitative MSE benchmarks, while the subsequent rows present reconstruction results from our PCDM, VAE-LMMSE, and PG-cGAN, respectively. 
\begin{itemize}
    \item Structural fidelity: The PCDM (Row 2) exhibits superior capability in recovering the sparse multipath structure. As shown in the amplitude and phase columns, PCDM accurately reconstructs the sharp peaks and distinct phase shifts, closely mirroring the ground truth. In contrast, the VAE-LMMSE (Row 3) suffers from the over-smoothing effect typical of Gaussian-based losses, resulting in blurred signal edges. The cGAN (Row 4), while sharper than VAE, introduces noticeable structural distortions and spurious noise pixels due to the instability of adversarial training.
    \item Error minimization: The amplitude/phase error heatmaps (Columns 3-4) further validate these findings. The error maps for PCDM are predominantly black, indicating near-zero residuals. Conversely, distinct bright spots are visible in the baselines' error maps, reflecting their inability to capture high-frequency channel variations.
    \item Quantitative superiority: The bar charts in the top-right corner eventually confirm the visual analysis. PCDM achieves the lowest MSE in both amplitude ($1.3 \times 10^{-21}$) and phase ($0.0326$), significantly outperforming the VAE-LMMSE ($3.5 \times 10^{-21} / 0.0670$) and cGAN ($1.8 \times 10^{-20} / 0.0786$).
\end{itemize}

\subsection{Performance Comparison in the Traffic Digital Twin}
To validate the effectiveness of the proposed orbit-adaptive ST-GNN in predicting LEO traffic, we compare it against three representative deep learning baselines, which involve adaptive graph convolutions and global attention mechanisms. The specific configurations are summarized in Table \ref{tab:traffic_baselines}.
\begin{itemize}
    \item Adaptive graph convolutional recurrent network (AGCRN)~\cite{bai2020adaptive}: It learns node-specific patterns through learnable node embeddings and generates an adaptive adjacency matrix. In our experiment, we feed it both traffic residuals and coordinate features to allow it to implicitly learn spatial dependencies.
    \item Dynamic spatiotemporal-aware GNN (DSTAGNN)~\cite{lan2022dstagnn}: This model constructs a dynamic graph based on the similarity of the input data itself. It utilizes a self-attention mechanism on the traffic data of the last time step to compute a dynamic adjacency matrix at every instance. Unlike our approach that uses physical coordinates, DSTAGNN infers topology purely from data correlations.
    \item Linear transformer (L-Transformer)~\cite{li2024linwa}: To capture long-range temporal dependencies without the quadratic complexity of standard transformers, we leverage a linear transformer. It utilizes a kernel-based linear attention mechanism to achieve $O(N)$ complexity. It is trained with a learning rate of $3 \times 10^{-3}$ and 2 attention heads.
    
\end{itemize}

\begin{figure}[!t]
    \centering
    \includegraphics[width=\linewidth]{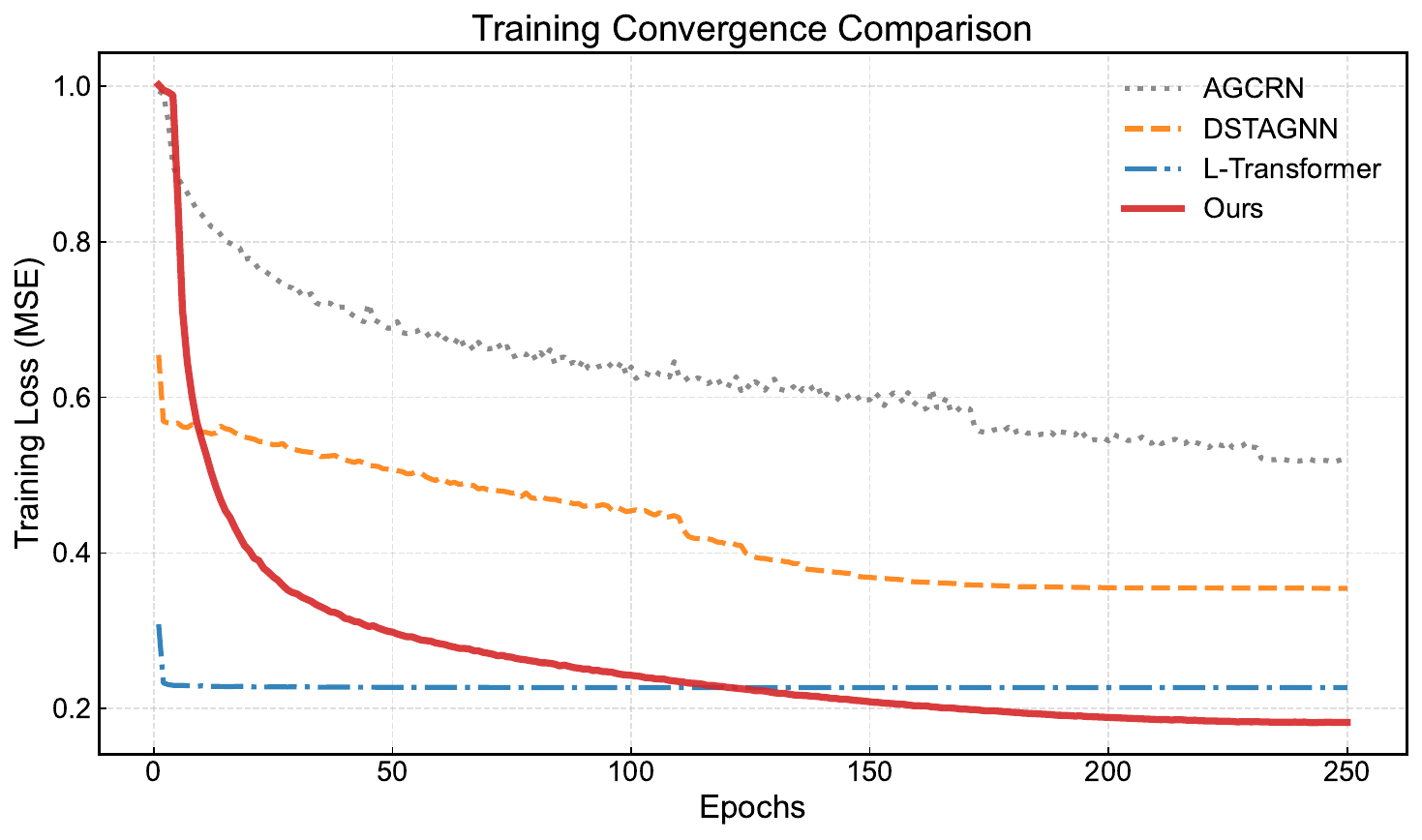}
    \caption{Training convergence comparison.}
    \label{fig:traffic_convergence}
\end{figure}

\begin{table}[!t]
\centering
\caption{Baseline Configuration in the Traffic Digital Twin}
\label{tab:traffic_baselines}
\begin{tabular}{l|ccc}
\toprule
\textbf{Model} & \textbf{AGCRN} & \textbf{DSTAGNN} & \textbf{L-Transformer} \\
\midrule
\multicolumn{4}{l}{\textit{A. Architecture Specifications}} \\
\midrule
Core Mech. & GCN+GRU & Attention+GRU & Linear attention \\
Graph Const. & Embedding & Similarity & N/A \\
Input Feat. & Traffic+Coord. & Traffic & Traffic \\
Hidden Dim. & 128 & 128 & 128 \\
\midrule
\multicolumn{4}{l}{\textit{B. Training Hyperparameters}} \\
\midrule
Optimizer & Adam & Adam & Adam \\
Learning Rate & $3 \times 10^{-3}$ & $5 \times 10^{-3}$ & $3 \times 10^{-3}$ \\
Batch Size & 32 & 32 & 32 \\
\bottomrule
\end{tabular}
\end{table}
To evaluate the learning stability and convergence speed of the proposed model in the traffic DT, we visualize the evolution of MSE loss during the training phase in Fig. \ref{fig:traffic_convergence}. The linear transformer (blue curve) exhibits a rapid descent within the first 10 epochs. This is attributed to its global attention mechanism, which quickly captures the dominant periodic patterns of satellite orbits. However, the curve becomes flat very quickly and remains at an MSE of approximately 0.23, which indicates its limitation in capturing fine-grained local topological changes. In contrast, the proposed orbit-adaptive ST-GNN (red curve) demonstrates a continuous optimization trajectory. Although the initial learning phase is slower due to the complexity of constructing dynamic spherical graphs, the model effectively surpasses the transformer after epoch 120 and achieves the lowest asymptotic error. This reflects that the dynamic graph structure progressively refines its understanding of the spatiotemporal dependencies.

\begin{figure}[!t]
    \centering
    \includegraphics[width=\linewidth]{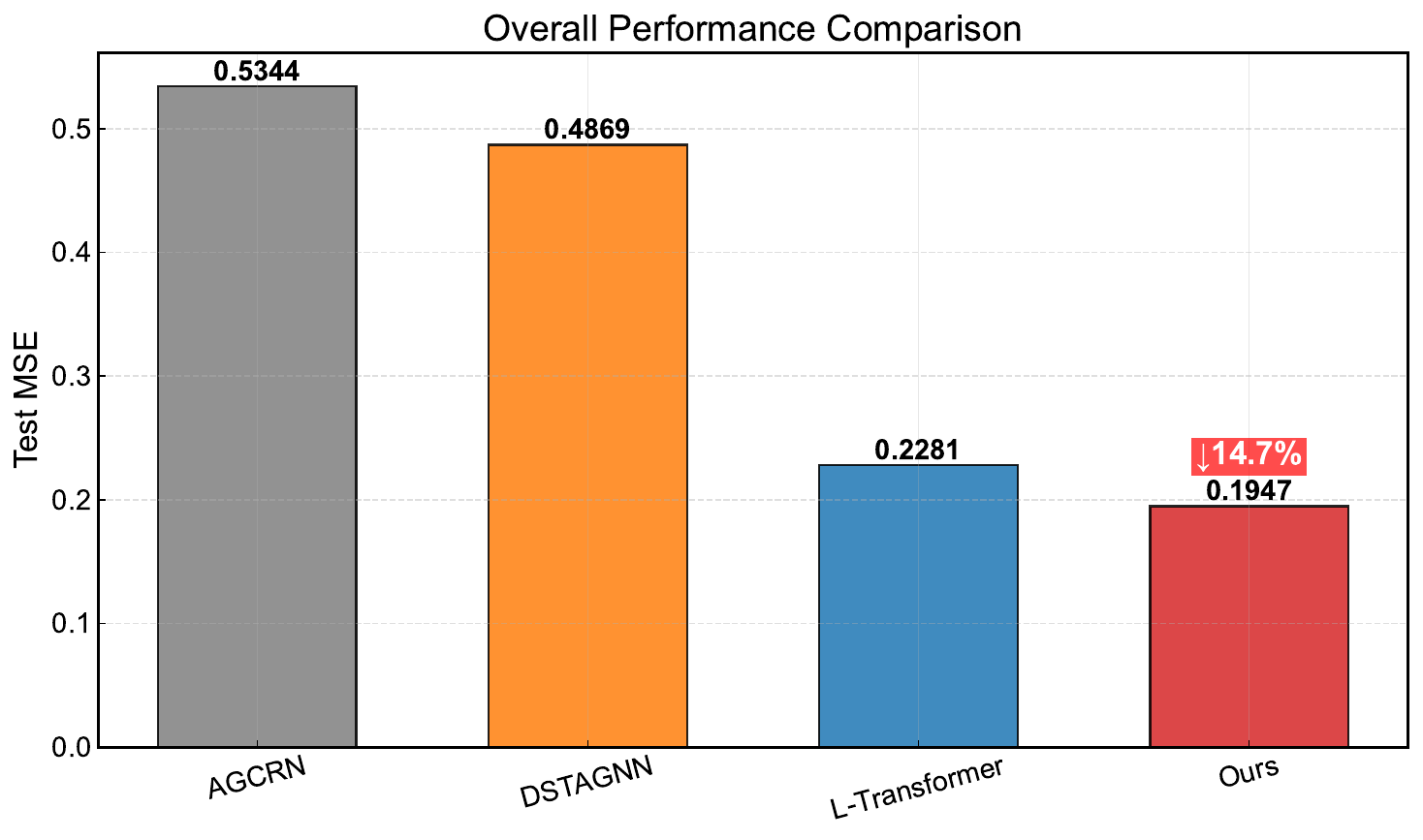}
    \caption{Overall performance comparison on the test dataset.}
    \label{fig:traffic_barchart}
\end{figure}

We further quantify the generalization capability of the trained models on the unseen test dataset. Fig. \ref{fig:traffic_barchart} presents the final test MSE for all baselines. The static graph approach (AGCRN) and the data-driven dynamic graph approach (DSTAGNN) yield relatively high errors (MSE > 0.48), which reflects that neither static embeddings nor purely data-driven correlations are sufficient for the highly dynamic LEO environment. The proposed orbit-adaptive ST-GNN achieves the lowest MSE of 0.1947. Compared to the second-best baseline (L-Transformer, MSE=0.2281), our model delivers a significant performance gain of 14.7\%. This clear gap demonstrates that using satellite trajectory data is much more effective than relying solely on historical traffic data.

\section{Conclusion}
\label{sec:Conclusion}
In this paper, we have proposed a physics-informed DT framework to address the challenges of channel aging, pilot sparsity, and traffic heterogeneity in NTN. By embedding deterministic physical laws into generative models, our framework ensures physical consistency in channel estimation and traffic prediction. The proposed framework integrates a channel DT based on physics-guided diffusion with a traffic DT based on orbit-adaptive residual learning, which realizes accurate state estimation in NTN.
The proposed framework offers a general paradigm for other high-mobility cyber-physical systems, such as unmanned aerial vehicle swarms and high-speed railways, characterized by the coexistence of rapid state evolution, sparse observations, and complex environmental heterogeneity. For the future work, we will investigate how NTN DTs can facilitate automated resource management.

%% *************************************************************************
%\section*{Acknowledgment}
%\addcontentsline{toc}{section}{Acknowledgment}
%
%\blindtext
%% *************************************************************************
%% References section
%%
%% Can use a bibliography generated by BibTeX as a .bbl file.
%%
\bibliographystyle{IEEEtran}
%% Argument is your BibTeX string definitions and bibliography database(s).
\bibliography{IEEEabrv,Ref}
%%
%% <OR> Manually copy in the resultant .bbl file.
%% Set second argument of \begin to the number of references.
%% (used to reserve space for the reference number labels box)
%%
%\begin{thebibliography}{1}
%	
%	\bibitem{IEEEhowto:kopka}
%	H.~Kopka and P.~W. Daly, \emph{A Guide to {\LaTeX}}, 3rd~ed.\hskip 1em plus
%	0.5em minus 0.4em\relax Harlow, England: Addison-Wesley, 1999.
%	
%\end{thebibliography}
%%
%% *************************************************************************
%\begin{IEEEbiography}[{\includegraphics[width=1in,height=1.25in,clip,keepaspectratio]{author1}}]{Author 1}
% \begin{IEEEbiography}{Author 1}
% \blindtext
% \end{IEEEbiography}

% \begin{IEEEbiography}{Author 2}
% \blindtext
% \end{IEEEbiography}
%% *************************************************************************
% \section*{Acronyms}
% \acuseall
% % \setlength{\mylabelwidth}{0.2\columnwidth}
% \IEEEprintacronyms
%% End All
\end{document}